\documentclass[11pt]{article}
\usepackage{amsmath}
\usepackage{amsfonts}
\usepackage{amssymb}
\usepackage[pdftex]{color,graphicx}
\usepackage{tikz}
\usetikzlibrary{matrix,arrows,decorations.pathmorphing}
\usepackage{tikz-cd}
\usetikzlibrary{arrows}

\newtheorem{Theorem}{Theorem}

\newtheorem{Cor}{Corollary}

\newtheorem{Lemma}{Lemma} 

\newtheorem{Def}{Definition}

\begin{document}

% Blackboard bold letters
\def\Z{\mathbb{Z}}
\def\R{\mathbb{R}}
\def\C{\mathbb{C}}
\def\S{\mathbb{S}}
\def\T{\mathbb{T}}
\def\H{\mathbb{H}}
\def\h{\mathbb{h}}

\newcommand{\ket}[1]{|{#1}\rangle}
\newcommand{\bra}[1]{\langle{#1}|}
\newcommand{\states}{\mathcal{S}}

\newcommand{\set}[1]{\ensuremath{ \lbrace #1 \rbrace }}
\newcommand{\Tr}{\text{Tr}}
\newcommand{\Span}[1]{\ensuremath{ \langle #1 \rangle }}
\newcommand{\ttd}{{\tt{d}}}

\def\pP{\mathcal{P}}
\def\cC{\mathcal{C}}
\def\oO{\mathcal{O}}
\def\iI{\mathcal{I}}
\def\eE{\mathcal{E}}
\def\sm{\mathfrak{s}}

\newcounter{expl}
\setcounter{expl}{2}

\title{Cohomological framework for contextual quantum computations}

\author{Robert Raussendorf\vspace{4mm}\\
{\small{\em{Department of Physics \& Astronomy, University of British Columbia, Vancouver, Canada,}}}\\
{\small{\em{Stewart Blusson Quantum Matter Institute, University of British Columbia, Vancouver, Canada}}}
}

\maketitle

\begin{abstract}We describe a cohomological framework for measurement based quantum computation, in which symmetry plays a central role. Therein, the essential information about the computational output is contained in topological invariants, namely elements of two cohomology groups. One of those invariants applies to the deterministic case, and the other to the general probabilistic case. The same invariants also witness quantumness in the form of contextuality. In result, they give rise to fundamental algebraic structures underlying quantum computation.
\end{abstract}

\section{Introduction}\label{Intro}

The Boolean algebra \cite{Boole} is at the foundation of all digital classical computation. From a quantum perspective it is thus pertinent to ask what its counterpart in quantum computation is.  {\em{Which fundamental algebraic structures can quantum computation be based on?}} This is the question the present paper is concerned with. We address it for the model of measurement-based quantum computation (MBQC) \cite{RB01}, in which the process of computation is driven by measurements rather than unitary evolution. While conceptually different, MBQC is equivalent to the circuit model in computational power.

We consider computational processes that start  with a classical input and end with a classical output; only the processing in-between is quantum. Shor's algorithm \cite{Shor} is an example: the number to be factored is a classical object, and so are its prime factors. Further we allow the classical input to vary, and the quantum computations in question thus evaluate functions.

Prototypical examples of algebraic structures describing aspects of quantum computation, or, more generally, of quantum mechanics, are the stabilizer formalism \cite{GoCha}--\cite{BVLN}, and quantum logic \cite{BvN}, \cite{ID}. In the computational setting of interest, we regard as minimal requirements on any algebraic structure $\mathfrak{S}$ underlying quantum computation that
\begin{itemize}
\item[(I)]{$\mathfrak{S}$ contains the function computed,}
\item[(II)]{$\mathfrak{S}$ contains a witness of quantumness.}
\end{itemize}
We identify two such structures, one applying to deterministic measurement-based quantum computations, and the other to probabilistic ones. The witness of quantumness required by Condition (II) arises as a contextuality witness.  Contextuality \cite{KS}--\cite{CSW}, also see \cite{GTS}--\cite{Coho2}, is a fundamental property of quantum mechanics, discovered in the wake of the Einstein-Podolsky-Rosen paradox \cite{EPR}. The statement ``quantum mechanics is contextual" means that descriptions of quantum phenomena in terms of classical statistical mechanics---so-called non-contextual hidden variable models (ncHVMs)---are in general not viable.  In such models, all observables are assigned pre-existing values which are merely revealed by measurement---in stark contrast to the quantum mechanical description.  Contextuality distinguishes quantum physics from classical physics.

The phenomenology explored in this paper is the triangle spanned by contextuality, MBQC, and cohomology plus symmetry.  The connection between MBQC \cite{RB01} and contextuality has previously been discussed in \cite{AB}--\cite{CF}. The link between contextuality and cohomology has been described in \cite{ABsheaf}, \cite{A2}, and in the formalism used here, which includes symmetry, in \cite{Coho}, \cite{Coho2}. 

This paper focusses on the remaining link of the triangle, between cohomology and symmetry on one side, and measurement-based quantum computation on the other. As we demonstrate below, in MBQC the essential information about the function computed and the presence of contextuality is cohomological. We identify two algebraic structures $\mathfrak{S}$ satisfying the above conditions (I) and (II), and then describe them in terms of cohomology. Symmetry and cohomology are united in the general probabilistic scenario, which is described by group cohomology.\medskip

To illustrate what these algebraic structures accomplish, we may contrast them with an existing result on contextuality in MBQC. Namely, the following holds \cite{RR13}.
\begin{Theorem}\label{NLPCrel}
Be ${\cal{M}}$ an MBQC with mod 2-linear classical processing relations, evaluating a function $o:(\mathbb{Z}_2)^m \longrightarrow \mathbb{Z}_2$. Then, ${\cal{M}}$ is contextual if it succeeds with an average probability $p_S>1-d_H(o)/2^m$, where $d_H(o)$ is the Hamming distance of $o$ from the closest linear function.
\end{Theorem}
While this theorem provides thresholds for witnessing contextuality in MBQCs, it says nothing about whether MBQCs exceeding these threshold actually exist. But they do. The simplest known example is a 3-qubit MBQC repurposing Mermin's star \cite{Merm}, see \cite{AB}. It generalizes to an infinite family of examples based on Reed-Muller codes \cite{MWS}. Further sporadic instances of contextual MBQCs derive from counterexamples to the former LU-LC conjecture \cite{LULC}.

The purpose of this paper is to provide a general framework within which these examples, and those to be found in the future, can be discussed---in a terminology that helps separate the general features from the particulars. A present constraint is that the MBQCs in question are non-adaptive, i.e., the choice of measurement bases does not depend on measurement outcomes obtained earlier. The generalization to adaptive MBQCs remains for future work. 

\medskip
The remainder of this paper is organized as follows. In Section~\ref{SumSec}, we summarize the results of this paper. Section~\ref{BG} is a review of background material. In Section~\ref{HMBQC} we describe $H$-MBQC, a variant of measurement-based quantum computation in which symmetry plays an important role. The subsequent discussion applies to this computational model. In Sections~\ref{comp} and \ref{ProbComp} we develop the cohomological viewpoint of deterministic and probabilistic MBQC, respectively. They contain the main results of this paper. We conclude in Section~\ref{concl}.

\section{Summary of results}\label{SumSec}

We introduce a generalized notion of MBQC based on symmetry. Each such $H$-MBQC transforms under a characteristic symmetry group $H$, whose action preserves the outputted function $o$. We establish the following properties.
\begin{itemize}
\item{The input structure to each $H$-MBQC is a group $Q$ related to $H$. Denote by $N$ the normal subgroup $N$ of $H$, such that each $n\in N$ maps the measurable observables of the computation element-wise to themselves, up to sign. Then, it holds that $Q=H/N$ (Theorem~\ref{quot}).}
\item{The output function $o: Q\longrightarrow \mathbb{Z}_2$ is preserved under the action of $H$ (Theorem~\ref{Invar}).}
\item{In the deterministic case, the output function $o$ is determined by a 2-cocycle $\beta_\Psi$, and the action of $Q$ on it (Theorem~\ref{PC}). In the probabilistic case, the output function $o$ is determined up to an additive constant by the phase function $\Phi$, a 1-cocycle in group cohomology that derives from the action of $H$ on the set of measurable observables (Theorem~\ref{o_Inv}).} 
\end{itemize}
These findings address Condition (I). In regards to Condition (II), on contextuality witnesses, the results of \cite{Coho} hold for the deterministic case, and results similar to those of \cite{Coho2} can be established in the probabilistic case. Namely, in the deterministic case, the $H$-MBQC is contextual if $[\beta_\Psi]\neq 0$ (Theorem~\ref{Thm_beta}) and also if $[\Phi]\neq 0$ \cite{Coho}. In the probabilistic case, an $H$-MBQC is contextual if it succeeds with a probability higher than a threshold $p_\text{crit}(\Phi)$ (Theorem~\ref{Tctx}). 

The cohomological aspects thereof are the following. First, regarding contextuality, in the deterministic case the cohomological formulation of the contextuality condition is apparent. In the probabilistic case, the contextuality threshold $p_\text{crit}(\Phi)$ depends on the phase function $\Phi$ only through its cohomology class $[\Phi]$; hence it is a cohomological invariant (Theorem~\ref{Hamming2}).

The output function $o$ is not a cohomological invariant, but at least the ``relevant'' part of it is. Namely, in any given $H$-MBQC there are ``trivial'' re-orientations of the measurement devices  which make the computation neither easier nor harder to perform, but change the outputted function. The output functions obtainable from one another through such transformations are therefore grouped into equivalence classes. These equivalence classes are cohomological invariants (Corollaries \ref{Cor1}, \ref{Cor2}).

From the perspective of the Criteria (I) and (II), all relevant information about a given $H$-MBQC is contained in cohomological invariants $[\beta_\Psi]$ (together with the action of $Q$ on the chain complex on which $\beta_\Psi$ is defined) and $[\Phi]$,  for the deterministic and probabilistic scenario, respectively.

\section{Preliminaries}\label{BG}

In this section we review background material. Section~\ref{MBQC} summarizes standard MBQC \cite{RB01}, and also reviews the simplest example of a contextual MBQC \cite{AB}, based on Mermin's star \cite{Merm}. We will expand on this example throughout. Section~\ref{BGctx} provides a definition of ``non-contextual hidden variable model'' tailored to the present physical setting. Sections~\ref{C} and \ref{PhaFu} cover elements of the cohomology of chain complexes and of groups. We review the basic definitions of cohomology, and define the chain complex applicable to our physical setting.  This material already appeared in \cite{Coho}, \cite{Coho2}, and we include it here for convenience of reference. 

\subsection{Summary of $l2$-MBQC}\label{MBQC}

The original MBQC scheme \cite{RB01} with cluster states as the universal computational resource is sometimes also called $l2$-MBQC, to emphasize the linearity of the classical side processing. This section is a recollection of basic facts about $l2$-MBQC that we subsequently refer to; it is no review by any means. For reviews of measurement-based quantum computation in general, see e.g. \cite{RBB03}, \cite{CLN}, \cite{RW12}. For the classical side processing in $l2$-MBQC see \cite{RB02}, or \cite{RR13}, \cite{CF} for brief summaries thereof.

Each MBQC requires classical side-processing. The need for it arises in two places: measurement bases must be adapted according to the computational input and previously obtained measurement outcomes, and the computational result must be extracted from those outcomes.  

In greater detail, $l2$-MBQC is characterized by the following properties:
\begin{enumerate}
\item{\label{Loc}The measurements are all local.}
\item{\label{CPR}For each measurement $i$, there are two possible choices for the measured observable $O_i[q_i]$, depending on a binary number $q_i$. Furthermore, both the bitwise output $\textbf{o}=(o_1,o_2..,o_k)$ and the choice of measurement bases, $\textbf{q}=(q_1,q_2,..,q_N)$ are functions of the measurement outcomes $\textbf{s}=(s_1,s_2,..,s_N)$. In addition, $\textbf{q}$ is also a function of the classical input $\textbf{i}=(i_1,i_2,..,i_m)$. These functional relations are all mod 2 linear,
\begin{subequations}\label{CCR}
\begin{align}\label{CCR_out}
\textbf{o}&=Z\textbf{s} \mod 2,\\ 
\label{CCR_in}
\textbf{q} &=T\textbf{s}+S\textbf{i} \mod 2.
\end{align}
\end{subequations}
Therein, the binary matrix $T$ encodes the temporal order in a given MBQC. If $T_{ij}=1$ then the basis for the measurement $i$ depends on the outcome of measurement $j$, hence the measurement $j$ must be executed before the measurement $i$.}
\end{enumerate}
For this MBQC setting, Theorem~\ref{NLPCrel} provides the connection with contextuality. We remark that without the mod 2-linear classical processing relations, e.g. for more than two choices of basis per measurement, the implication ``non-linearity $\Longrightarrow$ contextuality'' does not generally  hold \cite{FrembsQudit}. \medskip

The connection between contextuality and MBQC was first recognized in \cite{AB}. Therein it was demonstrated that the state-dependent version of Mermin's star \cite{Merm}---one of the simplest known proofs of quantum contextuality---can be repurposed as a small measurement-based quantum computation.
To motivate the question that will form the technical starting point for our investigation, we give a brief summary of the specific MBQC \cite{AB} related to Mermin's star (GHZ-MBQC).  We will return to this example throughout.

\paragraph{Reference example (GHZ-MBQC).} In this scenario, the resource state is a Greenberger-Horne-Zeilinger state $|\text{GHZ}\rangle= (|000\rangle + |111\rangle)/\sqrt{2}$  \cite{GHZ}, and the local measurable observables $O_i[q_i]$, depending on a binary number $q_i$, are $O_i[0]=X_i, \; O_i[1]=Y_i$, for $i=1,..,3$. The measurement outcomes $s_i \in \mathbb{Z}_2$ are related to the measured eigenvalues $\lambda_i = \pm 1$ of the respective local Pauli observables via $\lambda_i=(-1)^{s_i}$. There are two bits $a,b$ of input and one bit $o$ of output, and the computed function is an OR-gate, $o= a \vee b$. 

The required linear classical side processing is as follows. 
\begin{subequations}\label{CCRghz}
\begin{align}
\label{CCR_inGHZ}
q_1 = a,\, q_2 = b,\, q_3 = a+b \mod 2,\\
\label{CCR_outGHZ}
o= s_1+s_2+s_3 \mod 2.
\end{align}
\end{subequations}
The two input bits $a$ and $b$ determine the choices $q_i$ of measured observables through Eq.~(\ref{CCR_inGHZ}), and then the corresponding binary measurement outcomes $s_1, s_2, s_3$ determine the outputted value of the function, $o(a,b)$. 

Let's verify that the output is the intended OR function. First, consider $a=b=0$. Thus, by Eq.~(\ref{CCR_inGHZ}), $q_1=q_2=q_3=0$, and all three locally measured observables are of $X$-type. While the outcomes $s_1,s_2,s_3$ are individually random, they are correlated since the product of the corresponding observables $X_i$ is the stabilizer of the GHZ state, $X_1X_2X_3|\text{GHZ}\rangle = |\text{GHZ}\rangle$. Therefore, $s_1+s_2+s_3\mod 2 = 0$. Hence, with Eq.~(\ref{CCR_outGHZ}), $o(0,0)=0$ as required for the OR-gate.

We consider one more input combination, $a=0$ and $b=1$. Then, with Eq.~(\ref{CCR_inGHZ}), $q_1=0$ and $q_2=q_3=1$. Hence $X_1$, $Y_2$ and $Y_3$ are measured. Because of the stabilizer relation $X_1Y_2Y_3|\text{GHZ}\rangle = - |\text{GHZ}\rangle$, the three measurement outcomes $s_1,s_2,s_3$ satisfy $s_1+s_2+s_3 \mod 2 =1$. With Eq.~(\ref{CCR_outGHZ}), $o(0,1)=1$ as required. The discussion of the other two inputs is analogous.\smallskip

The OR-gate is a very simple function; yet it is of consequence for the above computational setting. Every MBQC requires a classical control computer, to enact the classical side processing of Eq.~(\ref{CCRghz}). This control computer is constrained to performing addition mod 2, and it is therefore not classically computationally universal. The OR-gate is a non-linear Boolean function. By adding it to the available operations, the extremely limited classical control computer is boosted to classical computational universality \cite{AB}.
\medskip

We now examine the above example for clues to help answer the initial question. For this purpose, it is useful to state Eq.~(\ref{CCR_outGHZ}) separately for all four input values.
\begin{equation}\label{CCRexpand}
\begin{array}{rrcr}
\textbf{input:} \; (0,0)& \hspace*{4mm}\textbf{output:} \; 0&=& s(X_1)+s(X_2)+s(X_3)\\
(0,1)& 1&=& s(X_1)+s(Y_2)+s(Y_3)\\
(1,0)& 1&=& s(Y_1)+s(X_2)+s(Y_3)\\
(1,1)& 1&=& s(Y_1)+s(Y_2)+s(X_3)
\end{array}
\end{equation}
Therein, if we assume the validity of a non-contextual hidden variable model describing the MBQC and regard the measurement outcomes $s(X_1),...,s(Y_3)$ as non-contextual  value assignments, then the four output relations yield Mermin's contradiction \cite{Merm}. Namely, adding the relations mod 2 gives $1=0$.

However, if in those relations we regard the values $s(X_1),...,s(Y_3)$ as quantum mechanical measurement record, the contradiction disappears. The relations then govern the classical side-processing in MBQC. We thus find that, for MBQC, contextuality and computation hinge on the same algebraic structure. If we impose an ncHVM description on top of this structure, we obtain a contradiction; and if we do not impose it, we obtain a computation.  Dropping an assumption puts a contradiction to work! This begs the question:  {\em{What precisely is this common algebraic structure underlying both parity-based contextuality proofs and measurement-based quantum computation?}}  

This question provides a starting point for our technical discussion. The answer hiding in Eq.~(\ref{CCRexpand}) will be presented first in Section~\ref{C}---for this particular example, as well as its generalizations.

\subsection{Contextuality}\label{BGctx}

To describe contextuality, we begin by defining its opposite---non-contextuality. 
We consider density matrices $\rho$, and the set of observables ${\cal{I}}$, which comprises all observables relevant for a given MBQC. This definition is from \cite{QuWi} where it has been applied to quantum computation with magic states.
\begin{Def}\label{def_nc_hvm}
A non-contextual hidden variable model (ncHVM) is a triple $(\Omega,q_{\rho}, \Lambda)$, with $q_{\rho}$ a probability distribution over a set $\Omega$ of internal states. The set $\Lambda=\{\lambda_{\nu}\}_{\nu \in \Omega}$ consists of functions, $\lambda_{\nu}: \iI \rightarrow \mathbb{C}$ obeying the following constraints:
\begin{enumerate}
\item For any set $S\subset  \iI$ of jointly measurable observables, there exists a quantum state $\ket{\psi}$ with the property
\begin{equation}\label{eq:qconsis}
X\ket{\psi}= \lambda_{\nu}(X) \ket{\psi}, \forall X \in S.
\end{equation}
\item The distribution $q_{p}$ satisfies:
\begin{equation}\label{eq:trq}
\text{tr}(A \rho)= \sum_{\nu \in \Omega} \lambda_{\nu}(A) q_{\rho}(\nu), \;\;\forall A \in \iI.
\end{equation}
\end{enumerate}
\end{Def}
For the present discussion, we always assume the ncHVM to be maximal, i.e., the set $\Omega$ of states is such that if $\nu$ satisfies Eq.~(\ref{eq:qconsis}) then $\nu \in \Omega$.

From condition (\ref{eq:qconsis}) it follows that for any triple $S=\{A,B,AB\}$ of jointly measurable observables, the functions $\lambda_{\nu}$ obey
\begin{equation}\label{LambdaConstr}
\lambda_{\nu}(AB)= \lambda_{\nu}(A) \lambda_{\nu}(B).
\end{equation}
With Definition~\ref{def_nc_hvm} in place, we can now define contextuality. A setting $(\rho,\iI)$ is contextual if it cannot be described by a non-contextual hidden variable model.

\subsection{Cohomology of chain complexes}\label{C}

Cohomology is useful for the present discussion because it provides a succinct characterization of parity-based contextuality proofs; see Lemma~\ref{BetaProp} below.

The basis for the cohomological description of the present phenomenology is a chain complex $\cC(E)$. It consists of one vertex, and edges, faces and volumes. Physically, the edges represent observables, the faces consistency constraints  among jointly measurable/inferable observables, and the volumes dependencies among the consistency constraints. 

For the edges exists a special notion of composition, ``$\oplus$'', whose physical motivation is a method of inference---or indirect measurement. We will describe this method of inference in greater detail in Section~\ref{IR}. As the definition of the chain complex below will depend on this composition, for now we may consider the operation ``$\oplus$'' as a free parameter. 

To build intuition, we describe here the notion of ``$\oplus$'' used in \cite{Coho}, \cite{Coho2} which is closely related. We denote the set of observables of interest in a given MBQC by ${\cal{I}}$, with the property that $A\in {\cal{I}} \Longleftrightarrow -A \in {\cal{I}}$. We denote by $E$ the set of equivalence classes $\{A,-A\}$ of observables, and define a map $\eta$ that picks a representative in each equivalence class,
\begin{equation}\label{etaDef}
E \ni a \mapsto \eta(a) = T_a \in {\cal{I}}.
\end{equation}
The set ${\cal{I}}$ can thus be written as 
\begin{equation}\label{DefE}
{\cal{I}}:=\{\pm T_a,\, a\in E\}.
\end{equation}
Now consider two commuting observables, $T_a$ and $T_b$. Then, the addition $a\oplus b$ of the edges $a,b \in E$ is defined via 
$$
T_{a\oplus b} = \pm T_aT_b.
$$ 
We reiterate that the definition of the set ${\cal{I}}$ and of the composition operation ``$\oplus$'' will be refined in Section~\ref{IR}.  Specifically, the change we will make to the above definition of the composition $a\oplus b$ is in restricting $a$ to a subset of $E$. The physics informing these modifications is not background material, and the full definition is  postponed until after this physics has been discussed.

Specifying the sign factor in in the above relation, we have
\begin{equation}\label{prod_T}
T_{a\oplus b} = (-1)^{\beta(a,b)} T_aT_b.
\end{equation}
The function $\beta$ features prominently in parity based contextuality proofs, as we illustrate below.

\begin{figure}
\begin{center}
\begin{tabular}{lclcl}
(a) && (b) && (c)\\
\includegraphics[height=3.5cm]{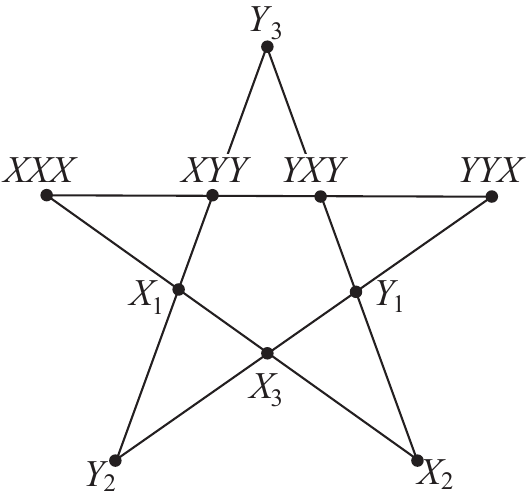} &&
\includegraphics[height=3.3cm]{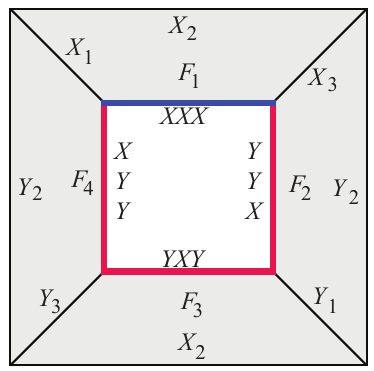}  &&
\includegraphics[height=3.3cm]{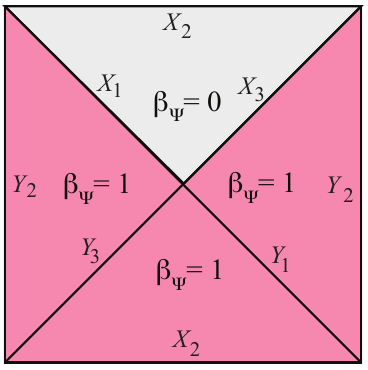} 
\end{tabular}
\caption{\label{MermSt} Mermin's star. (a) Standard representation. Each line represents a measurement context, composed of four commuting Pauli observables multiplying to $\pm I$. (b) Mermin's star re-arranged on a surface. The Pauli observables now correspond to edges, and each measurement context to the boundary of one of the four elementary faces. The exterior edges are pairwise identified. The colored edges carry a value assignment, resulting from the GHZ stabilizer. (c) Relative complex ${\cal{C}}(E,E_0)$. The edges corresponding to observables in the GHZ stabilizer are removed by contraction.}
\end{center}
\end{figure}

With the above definitions, the complex $\cC(E)$ is constructed as follows.
\begin{enumerate}
\item $C_0(E)=\mathbb{Z}_2$. Geometrically, we have a single vertex.
\item $C_1(E)$ is freely generated as a $\mathbb{Z}_2$-module by the elements $[a]$ where $a\in E$. These labels correspond to the set of edges.
\item $C_2(E)$ is freely generated as a $\mathbb{Z}_2$-module by the pairs $[a|b]$ for which $a\oplus b$ is defined. The pairs $(a,b)$ correspond to faces. We denote the set of all faces by $F$.
\item $C_3(E)$ is freely generated as a $\mathbb{Z}_2$-module by the triples $[a|b|c]$ where $a,b,c\in E$ pair-wise commute and the labels $a\oplus b$, $b\oplus c$, and $a\oplus b \oplus c$ are defined. These triples $(a,b,c)$ correspond to volumes and the set of volumes will be denoted by $V$.
\end{enumerate} 
In our applications, the physical information lives on the edges, faces and volumes of $\cC(R)$, not on the vertices. Wlog. we may therefore merge all vertices into one. In Figs. \ref{MermSt} b,c and \ref{lemon}, the endpoints of edges are not identified, for better graphical display. 
\medskip

The differentials in the complex
$
C_3(E)\stackrel{\partial}{\rightarrow} C_2(E) \stackrel{\partial}{\rightarrow} C_1(E) \stackrel{\partial}{\rightarrow} C_0(E)
$
are defined as 
\begin{equation}\label{partial}
\partial[a]=0,\;\; \partial[a|b]=[b]-[a\oplus b]+[a],\;\; \partial[a|b|c]=[b|c]-[a\oplus b|c]+[a|b\oplus c]-[a|b].
\end{equation}
 Given the chain complex $\cC_*(E)$, there is a corresponding cochain complex $\cC^*(E)$ as usual.
 The cochains $C^n(E)$ are $\Z_2$-linear maps $C_n(E)\rightarrow \Z_2$. Equivalently we can think of the cochains as functions on the basis elements of $C_n(E)$.  The abelian group structure on the cochains is represented by addition of functions.  
 
The coboundary operator $d:C^n(E)\rightarrow C^{n+1}(E)$ is defined by 
$$
dc(x)=c(\partial(x)),
$$ 
where $c\in C^n(E)$ and $x\in C_{n+1}(E)$.\medskip

The function $\beta$ defined in Eq.~(\ref{prod_T}) has a cohomological interpretation. Namely, $\beta$ is a 2-cocycle in $\cC(E)$. It is a 2-cochain by definition, and 
\begin{equation}\label{betaCyc}
d\beta=0
\end{equation} 
follows from associativity of operator multiplication, $(T_aT_b)T_c=T_a(T_bT_c)$, together with Eq.~(\ref{prod_T}). Then,
$$
[\beta] =\{ \beta + dx,\; \forall x\in C^1(\cC(E))\}
$$
is the cohomology class of the cocycle $\beta$. The possible cohomology classes $[\beta]$ form the second cohomology group $H^2(\cC(E),\mathbb{Z}_2)$ of $\cC(E)$.\smallskip

A first demonstration of the efficiency of the topological calculus is the following result \cite{Coho}.

\begin{Lemma}\label{BetaProp}
The set ${\cal{I}}$ of observables contains a parity-based contextuality proof if and only if $[\beta]\neq 0$.
\end{Lemma}
This result is illustrative of the use of cohomology in the description of contextuality, and therefore we reproduce the proof here.\smallskip

{\em{Proof of Lemma~\ref{BetaProp}.}} Denote by $\mathfrak{s}: E \longrightarrow \mathbb{Z}_2$ an assumed non-contextual value assignment for the observables $\{T_a,\; a\in E\}$, with $(-1)^{\mathfrak{s}(a)}$ the measured eigenvalue for $T_a$. Then, for every pair $T_a,T_b$ of simultaneously measurable observables Eq.~(\ref{prod_T}) implies that
$$
\mathfrak{s}(a)+\mathfrak{s}(b)+\mathfrak{s}(a\oplus b) \mod 2 = \beta(a,b).
$$
Noting that $\mathfrak{s}$ is a 1-cochain in the complex $\cC(E)$, and for the face $(a,b)\in C_2(\cC(E))$ it holds that $\partial (a,b) = a + b + (a\oplus b)$, we can rewrite this in cohomological form as $d\mathfrak{s}(a,b) = \beta(a,b)$. The entire system of linear constraints representing the parity-based contextuality proof thus is
\begin{equation}\label{beta_ds}
d\mathfrak{s} = \beta.
\end{equation}
This system of equations has a solution $\mathfrak{s}$ if and only if $[\beta]=0$. The other way around, it has no solution for $\mathfrak{s}$, meaning it represents a parity-based contextuality proof, if and only if $[\beta]\neq 0$. $\Box$\medskip

Lemma~\ref{BetaProp} characterizes state-independent parity-based contextuality proofs in a cohomological fashion. However, here we are interested in {\em{state-dependent}} proofs, because MBQC uses a resource state. We therefore need to adapt the cohomological calculus to accommodate this state.

The corresponding topological characterization lives in a relative complex $\cC(E,E_0)$, derived from the complex $\cC(E)$ introduced above. In the deterministic case, the MBQC resource state enters the picture through the set $E_0 \subset E$. Namely, the resource state is an eigenstate of the observables $T_a$ for all $a\in E_0$. For now, we treat the set $E_0$ as a free parameter; the definition of $E_0$ suitable to the computational setting of interest, applicable to both deterministic and probabilistic computations, will be given in Section~\ref{iG}.

Geometrically, ${\cal{C}}(E,E_0)$ is created from $\cC(E)$ by collapsing the edges, faces, and volumes coming from $E_0$. Mathematically, $\cC(E,E_0)$ is constructed as follows. First, we construct the chain complex $\cC_*(E_0)$ for the subset $E_0$. The inclusion $E_0\subset E$ gives an inclusion of the chain complexes $\cC_*(E_0)\subset \cC_*(E)$. 
The relative complex $\cC_*(E,E_0)$ is defined as the quotient $\cC_*(E)/\cC_*(E_0)$ meaning that in each degree $C_n(E,E_0)$ is given by the quotient group $C_n(E)/C_n(E_0)$. The basis is obtained by erasing the basis elements of $C_n(E_0)$ from the basis elements of the larger complex $C_n(E)$.

The relative boundary operator $\partial_R$ is induced from the boundary operator $\partial$ of $\cC_*(E)$, and it can be calculated by applying $\partial$ and removing the chains which lie in $\cC_*(E_0)$.   The relative cochain complex $C^n(E,E_0)$ consists of cochains in $C^n(E)$ whose restriction to $C_n(E_0)$ is zero. The relative coboundary operator is the same as the coboundary operator of $\cC^*(E)$.
 
In the deterministic case, the observables $T_a$, for $a \in E_0$ have deterministic values $s_a$. With $|\Psi\rangle$ the $H$-MBQC resource state,
\begin{equation}\label{eval}
T_a |\Psi\rangle = (-1)^{s_\Psi(a)}|\Psi\rangle, \;\; \forall a\in E_0.
\end{equation}
Extending the function $s_\Psi$ to all of $E$ via $\overline{s}_\Psi|_{E_0} =s_\Psi$, $\overline{s}_\Psi(a)=0$ for all $a \in E\backslash E_0$, we define
\begin{equation}\label{betaPsiDef}
\beta_\Psi:= \beta + d\overline{s}_\Psi.
\end{equation}
The 2-cochain $\beta_\Psi$ is in fact a 2-cocycle in $\cC(E,E_0)$, since, with Eq.~(\ref{betaCyc}), $d\beta_\Psi = d \beta + d d \overline{s}_\Psi =0$.
 \medskip

{\em{Example, Part~\theexpl.}} We illustrate the procedure of contracting $\cC(E)$ to $\cC(E,E_0)$ with the GHZ-MBQC. Fig.~\ref{MermSt}b shows the complex $\cC(E)$, before the contraction, and Fig.~\ref{MermSt}c the complex $\cC(E,E_0)$ resulting from the contraction. The edges to be contracted are $a_{XXX}$, $a_{YYX}$, $a_{YXY}$ and $a_{XYY}$, labeling stabilizer elements of the GHZ resource state. 

Let us further check that the cohomological method does indeed flag the state-dependent GHZ scenario as contextual. There are four elementary faces in ${\cal{C}}(E,E_0)$, $F_1', .., F_4'$, which derive from the corresponding faces in ${\cal{C}}(E)$; see Figs.~\ref{MermSt} b, c. The edges contracted in $\cC(E,E_0)$ are $a_{XXX}$, $a_{YYX}$, $a_{YXY}$ and $a_{XYY}$, labeling stabilizer elements of the GHZ resource state. The corresponding values are $s(a_{XXX})=0$, $s(a_{YYX})=s(a_{YXY})=s(a_{XYY})=1$. Before the contraction, $\beta$ evaluates to zero on all four shaded faces shown in Fig.~\ref{MermSt}b, $\beta(F_1)=\beta(F_2)=\beta(F_3)=\beta(F_4)=0$. After the contraction, with Eq.~(\ref{betaPsiDef}), $\beta_\Psi$ evaluates on the corresponding faces in $\cC(E,E_0)$ to $\beta_\Psi(F_1')=0$, $\beta_\Psi(F_2')= \beta_\Psi(F_3')=\beta_\Psi(F_4')=1$; see Fig.~\ref{MermSt}c.

Denote $F'=F_1'+F_2'+F_3'+F_4'$ such that the relative boundary of $F'$ vanishes, $\partial_R F'=0$. Now assume that $[\beta_\Psi]=0$, i.e., $\beta_\Psi=ds$ for some non-contextual value assignment  $s\in C^1({\cal{C}}_R,\mathbb{Z}_2)$. Then, 
$$
1=\int_{F'}\beta_\Psi = \int_{F'}ds = \int_{\partial F'} s = 0.
$$ 
Contradiction. Hence, no non-contextual value assignment $s$ exists.

\subsection{Cohomology of groups}\label{PhaFu}

This section introduces elements of group cohomology that we will need in the present discussion. For a comprehensive introduction, see \cite{AM} or \cite{Weib}; for a summary of the essential ideas, see \cite{CfA}.

The symmetry group $H$ that we will fully introduce in Section~\ref{sG} has the property that it maps the set ${\cal{I}}$ of observables to itself. All transformations $h \in H$  can be written in terms of a {\em{phase function}} $\tilde{\Phi}$,
\begin{equation}\label{PhiTilDef}
h(T_a) = (-1)^{\tilde{\Phi}_h(a)}T_{ha}.
\end{equation}
This relation is simultaneously a definition of the phase function, and of the action of $H$ on $E$ induced by the action on ${\cal{I}}$. The phase function is an important object for the present discussion. It has two arguments, an edge $a \in E$ and a group element $h \in H$, and it takes values in $\mathbb{Z}_2$. We remark that the phase function has previously been used in physics, namely in the subject of Fourier space crystallography \cite{Merm2}. The connection to group cohomology was established in \cite{CfA}.\medskip

To further build the cohomological calculus, the above definition makes $\tilde{\Phi}$ a 1-cochain in group cohomology. (We sometimes use a slightly different notation for the phase function to emphasize this fact, namely $\tilde{\Phi}(h,a)$ instead of $\tilde{\Phi}_{h}(a)$.) Group cohomology is based on a coboundary operator ${\tt{d}}$, whose action on the 1-cochain $\tilde{\Phi}$ is defined to be
\begin{equation}\label{CG2}
 \ttd \tilde{\Phi}(h_1,h_2,a)= \tilde{\Phi}(h_1,h_2a)+\tilde{\Phi}(h_1h_2,a)+\tilde{\Phi}(h_2,a).
\end{equation}
We require $\tilde{\Phi}$ to be compatible with the action of a group on ${\cal{O}}$, i.e., $(h_1h_2)(T_a)=h_1(h_2(T_a))$ for all $h_1,h_2\in H$. With Eq.~(\ref{PhiTilDef}) it follows that
$$
\ttd \tilde{\Phi} = 0.
$$
The phase function is thus a 1-cocycle. 

For the 0-cochains (w.r.t. group cohomology) $\beta$ and $\chi$ we have, for $\mathbb{Z}_2$ coefficients,
\begin{equation}\label{CG1}
\ttd \chi(h,a) = \chi(ha)+\chi(a),\;\;\;\; \ttd\beta(h,f)=\beta(hf)+\beta(f).
\end{equation}
Then,
$$
[\tilde{\Phi}] =\{\tilde{\Phi}+{\tt{d}}\chi,\; \forall \chi\}
$$
is the cohomology class of the cocycle $\tilde{\Phi}$. It is an element of the first cohomology group of the group $H$.
\medskip

As a first illustration of the usefulness of the symmetry group $H$, the phase function $\tilde{\Phi}$ and the calculus of group cohomology  we show how a contextuality proof can be based on them \cite{Coho}. Assume that a consistent value assignment $s$ exists for the observables of from a set ${\cal{I}}$. Then, the transformation law Eq.~(\ref{PhiTilDef})  for observables  implies that a new value assignment $h\cdot s$ is generated from the old assignment $s$,
$$
h\cdot s (a) = \tilde{\Phi}_h(a) + s(ha),\; \forall h\in H.
$$
\stepcounter{expl}{\em{Example, Part~\theexpl.}} For the state-dependent Mermin star, we consider the transformation $h_0=A_1A_2Y_3 \in H$, which leaves the GHZ-state invariant. It holds that $\tilde{\Phi}_{h_0}(a_{X_3})=1$, and $\tilde{\Phi}_{h_0}(a_{Y_3})=\tilde{\Phi}_{h_0}(a_{X_1})=\tilde{\Phi}_{h_0}(a_{Y_1})=0$. We further have a partial value assignment $s_\Psi(a_{XXX})=0$, $s_\Psi(a_{XYY})=s_\Psi(a_{YXY})=s_\Psi(a_{YYX})=1$, and the same values hold for the assignment $h_0\cdot s_\Psi$.

Now assume that the partial value assignment $s_\Psi$ can be extended to a global non-contextual value assignment $s$. We then have \cite{Coho},
$$
\begin{array}{rcl}
1 &=& h_0\cdot s(a_{XXX})+h_0\cdot s(a_{YXY})\\
&=& h_0\cdot s(a_{X_1})+h_0\cdot s(a_{X_3})+ h_0\cdot s(a_{Y_1})+h_0\cdot s(a_{Y_3})\\
&=& s(a_{Y_1})+\tilde{\Phi}_{h_0}(a_{X_1}) + s(a_{X_3})+\tilde{\Phi}_{h_0}(a_{X_3}) + s(a_{X_1})+\tilde{\Phi}_{h_0}(a_{Y_1}) + s(a_{Y_3})+\tilde{\Phi}_{h_0}(a_{Y_3}) \\
&=& s(a_{X_1})+s(a_{X_3})+ s(a_{Y_1})+s(a_{Y_3}) + 1\\
&=& s(a_{XXX})+s(a_{YXY})+1\\
&=& 0+1+1 =0.
\end{array}
$$ 
Contradiction. No consistent value assignment $s$ is compatible with $s_\Psi$.\smallskip

Let's take a step back and analyze why the above proof worked. The key property of the phase function $\tilde{\Phi}$ is
\begin{equation}\label{Phi_NT}
\tilde{\Phi}_{h_0}(a_{X_1}) +\tilde{\Phi}_{h_0}(a_{X_3}) +\tilde{\Phi}_{h_0}(a_{Y_1}) +\tilde{\Phi}_{h_0}(a_{Y_3})  = 1.
\end{equation}
Eq.~(\ref{Phi_NT}) is an obstruction to $\tilde{\Phi}$ being a trivial cocycle. Denote the 1-chain $a_{X_1}+a_{X_3}+a_{Y_1}+a_{Y_3}=:b$. Then, we have the property that $h_0b=b$, and Eq.~(\ref{Phi_NT}) in cohomological notation reads $\tilde{\Phi}(h_0,b)=1$. Now assume that $\tilde{\Phi}={\tt{d}}x$. Then, ${\tt{d}}x(h_0,b) = x(h_0b) + x(b) =0$. Contradiction. Hence, $\tilde{\Phi} \neq {\tt{d}}x$, $\forall x$. For any symmetry group of the above setting containing $h_0$, $\tilde{\Phi}$ is a non-trivial group cocycle.

\section{A generalized notion of MBQC}\label{HMBQC}

In this section we formally introduce a generalization of standard MBQC in which the input is described by a group. We call this MBQC-variant $H$-MBQC, to emphasize the role of symmetry. It is for this computational model that we establish computational structures $\mathfrak{S}$, satisfying Criteria (I) and (II),  in Sections~\ref{comp} and \ref{ProbComp}.

Compared to the characteristics listed in Section~\ref{MBQC}, this leads to two relaxations.
\begin{itemize}
\item[\ref{Loc}.]{The measurements in MBQC are not required to be local or even pairwise commuting. We will enforce a weaker property, inferability, which is defined in Section~\ref{IR}. This notion of inference is reminiscent of syndrome measurement in subsystem codes; see e.g. \cite{Bac}, \cite{Bombin}.}
\item[\ref{CPR}.]{In Eq.~(\ref{CCR_in}), the vector space of input is replaced by a finite group, which may be Abelian or non-Abelian. In result, there can be more than two settings per measurement.}
\end{itemize}
On the other hand, in this paper we will be more restrictive than the standard scheme \cite{RB01} in one respect. Namely, we only discuss non-adaptive MBQCs ($T=0$ in Eq.~(\ref{CCR_in})). Adaptive MBQCs will have to be addressed in a future, more detailed treatment. We also restrict to a single bit of output, but only for notational simplicity. 

Below we describe three notions at the basis of $H$-MBQC, namely {\em{resolutions}} of observables with inferable outcomes, the {\em{input group}} and the {\em{symmetry group}}.  

\subsection{Resolutions of observables}\label{IR}

In MBQC, the outcomes of the measurements driving the computation are individually completely random, and therefore not of interest one-at-a-time. Of interest are only their correlations. By measuring individual observables and classically post-processing the outcomes, we {\em{infer}} values of other observables that are, a priori, harder to measure.

For example, in GHZ-MBQC, for the input (0,0), the observables $X_1$, $X_2$ and $X_3$ are measured. While the three outcomes are individually random, jointly they imply the value of the correlated observable $X_1X_2X_3$, which is deterministic and represents computational output.\medskip 

We now formalize this notion of inference. To this end, we first define the set ${\cal{O}}$, which is the set of physically measurable observables in a given MBQC. The set ${\cal{O}}$ has to be distinguished from the larger set ${\cal{I}}$ of observables ``of interest for MBQC'' as introduced in Section~\ref{C}. We will sharpen the definition of the set ${\cal{I}}$ below. Motivated by the computational setting~\cite{RB01}, each observable $A \in {\cal{O}}$ is constrained to have eigenvalues $\pm 1$ only\footnote{It is straightforward to extend the analysis to observables with eigenvalues $e^{i2\pi\, k/d}$, $k,d\in \mathbb{N}$.}, and furthermore, it holds that $A \in {\cal{O}} \Longleftrightarrow -A \in {\cal{O}}$.

We can now state the notion of joint measurability that we apply in this paper. 
\begin{Def}\label{JM}
Two observables $A$ and $B$ are jointly measurable if (i) $A$ is directly measurable, i.e. $A \in {\cal{O}}$, (ii) the value of $B$ can be inferred through sequential measurement of observables in ${\cal{O}}$, and classical processing, and (iii) $[A,B]=0$.
\end{Def}
Note that this is slightly stronger than commutativity, since the observable $A$ is restricted to the set ${\cal{O}}$ of directly measurable observables.

The physical motivation for this definition is the following procedure for joint measurement. First, measure $A$ directly (which is possible since $A \in {\cal{O}}$), then implement the procedure to infer the value of $B$ (which exists by assumption). Since $A$ commutes with $B$, measuring $A$ first does not affect the value for $B$.

We now  describe the method of inference applied in the present physical setting. The elementary step of inference follows from Def.~\ref{JM}. Namely, if $A$ and $B$ can be jointly measured, then the measured eigenvalue $\lambda(AB)$ of $AB$ can be inferred, 
\begin{equation}\label{3lamb}
\lambda(AB) = \lambda(A)\lambda(B).
\end{equation}
Thus, if $A$ and $B$ are jointly measurable, with $A$ being directly measurable and the value of $B$ being inferable through some measurement sequence, then the value of $AB$ is also inferable. This is the elementary step of inference. It leads to the following definition.
\begin{Def}\label{Infer}
Given an observable $X$, the resolution of $X$ in ${\cal{O}}$ is a product
\begin{equation}\label{res}
X =  A_1(A_2(A_3(...(A_{K-1}A_K)))),
\end{equation}
for some integer $K\geq1$. Therein, $A_i \in {\cal{O}}$, $\forall\, i=1,..,K$, and $B_i:= \prod_{l=i+1}^K A_l$, $\forall i=0,.., K-1$, such that $[A_i,B_i]=0$, $\forall i=1,..,K-1$. If $X$ has a resolution in ${\cal{O}}$ then it is called inferable. 
\end{Def}
The physical motivation for this definition is the following result.
\begin{Lemma}\label{InferMeth}
For any observable $X$ with a resolution in ${\cal{O}}$, the value of $X$ can be inferred through sequential measurement of observables in ${\cal{O}}$, and classical post-processing. 
\end{Lemma}
Lemma~\ref{InferMeth} gives physical meaning to the notion of ``inferable''. Namely ``$X$ is inferable'' means that the value of $X$ can be obtained through a particular sequence of measurements, captured by the resolution of $X$.\smallskip

{\em{Proof of Lemma~\ref{InferMeth}.}} The proof is by induction. $B_{K-1}=A_K \in {\cal{O}}$ is directly measurable. By the structure of the resolution, $[A_i,B_i]=0$, and hence $A_i$ and $B_i$ are jointly measurable if the value of $B_i$ can be inferred through measurements in ${\cal{O}}$. With the elementary step of inference, Eq.~(\ref{3lamb}), the value of $B_{i-1}=A_iB_i$ can be inferred if the value of $B_i$ can. By induction, the value of $B_0$ can be inferred through a sequence of measurements in ${\cal{O}}$; and $B_0=X$. $\Box$\medskip

Lemma~\ref{InferMeth} motivates the following sharpening of the definition of ${\cal{I}}$, the set of observables ``of interest'' in given MBQC.
\begin{Def}
${\cal{I}}$ is the set of all observables which have a resolution Eq.~(\ref{res}) in ${\cal{O}}$. 
\end{Def}
{\em{Remark:}} The above definitions of the set ${\cal{I}}$ and ${\cal{O}}$ and of the elementary step of inference are very similar to those used in \cite{QuWi}, providing a conceptual link between MBQC as discussed here, and quantum computation with magic states as discussed in \cite{QuWi}. The only difference is that in \cite{QuWi} the observables are constrained to be of Pauli type, while here they are not.\medskip

We have so far completed the definitions that are essential from the perspective of physics, namely of the sets ${\cal{O}}$ and ${\cal{I}}$, and of the resolution. We are now ready to state the main result of this section, namely how the value of an observable $X$ with resolution in ${\cal{O}}$ is inferred from the measurement outcomes of the observables in the resolution of $X$. Denoting the measured eigenvalues for the observables $A_i$ by $(-1)^{s(a_i)}$ and the inferred eigenvalue for $X$ by $(-1)^{s(x)}$, from Eq.~(\ref{res}) it follows that
\begin{equation}\label{out}
s(x) = \sum_{i=1}^K s(a_i) \mod 2.
\end{equation}
Eq.~(\ref{out}) is of exactly the same form as Eq.~(\ref{CCR_out}), relating MBQC measurement outcomes to the computational result. The notion of inference provided by Definition~\ref{Infer} thus describes the extraction of computational output from the measurement outcomes.\smallskip

\stepcounter{expl}{\em{Example, Part~\theexpl.}} With respect to the above definitions, for ${\cal{O}}=\{\pm X_i, \pm Y_i,\; i=1,2,3\}$, the state-independent Mermin star is {\em{not}} contextual (and that is a good thing). The inference according to Def.~\ref{Infer} implements the four Mermin constraints invoking local observables, i.e., $\mathfrak{s}(a_{X_1})+\mathfrak{s}(a_{X_2})+\mathfrak{s}(a_{X_3})=\mathfrak{s}(a_{XXX})$. But it does not provide the fifth relation, $\mathfrak{s}(a_{XXX})+\mathfrak{s}(a_{XYY})+\mathfrak{s}(a_{YXY})+\mathfrak{s}(a_{YYX})=1$, because the observables referred to therein are not jointly measurable in the given setting. As a consequence, ncHVM value assignments $\mathfrak{s}$ exist, and correspondingly the Mermin inequality can be written down. This is important for contextuality in MBQC, cf. Theorem~\ref{NLPCrel}.

We further remark that, while value assignments $\mathfrak{s}$ exist, none matches $s_\Psi$ defined in Eq.~(\ref{eval}) on $E_0=\{a_{XXX},a_{XYY},a_{YXY},a_{YYX}\}$; hence the current observation is compatible with the {\em{state-dependent}} cohomological contextuality proof described in Part 2 of this example.
\medskip 

In the remainder of this section we address a mathematical aspect pertaining to the definition of the chain complexes $\cC(E)$ and $\cC(E,E_0)$ introduced in Section~\ref{C}, in particular the operation ``$\oplus$'' for the composition of edges. This discussion had to wait for the definition of joint measurability and of the sets ${\cal{O}}$ and ${\cal{I}}$, as given above. Denoting $$E_{\cal{O}}:=\{a \in E|\, T_a \in {\cal{O}}\},$$ the composition $a\oplus b$ of two edges $a$ and $b$ is defined if and only if $a\in E_{\cal{O}}$, $b\in E$ and $[T_a,T_b]=0$. Specifically, $a\oplus b$ is defined through the relation
\begin{equation}\label{oplusDef}
T_{a\oplus b} = \pm T_aT_b, \;\;\forall a\in E_{\cal{O}},\,b\in E\;\text{s.th. }[T_a,T_b]=0.
\end{equation}
 This affects the chain complexes $\cC(E)$ and $\cC(E,E_0)$ introduced in Section~\ref{C} to model the present physical setting. Recall that in Section~\ref{C} we left the composition operation``$\oplus$'' as a free parameter. The law of composition is now fixed. For the remainder of this paper, Eq.~(\ref{oplusDef}) applies.

The physical relevance of Eq.~(\ref{oplusDef}) is that it represents an elementary step of inference at the level of chain complexes. Namely,  for every elementary step of inference, deducing the value of $T_{a\oplus b} \in {\cal{I}}$ from the values of $T_a\in {\cal{O}}$ and $T_b \in {\cal{I}}$, the complex $\cC(E)$ has a face $(a,b)$, with boundary $\partial (a,b) = a + b + (a\oplus b)$. \smallskip

\begin{figure}
\begin{center}
\includegraphics[width=5.5cm]{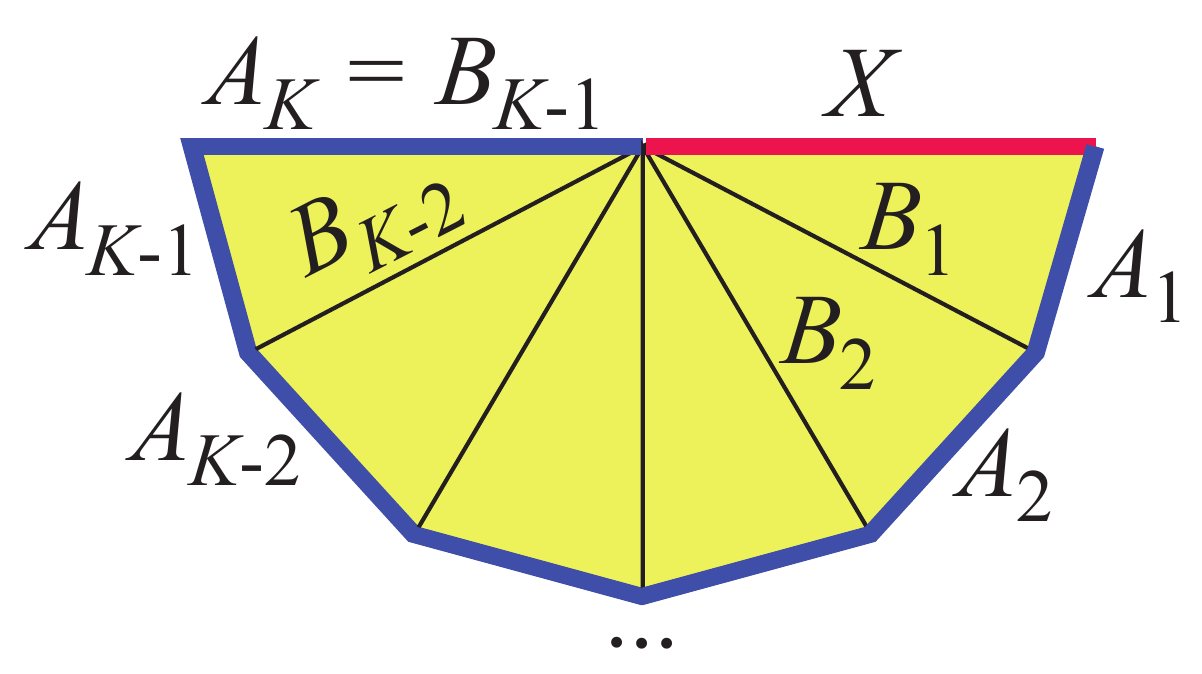}
\caption{\label{lemon}The resolution of an observable $X\in {\cal{I}}$, and corresponding surface $f\in C_2(E)$.}
\end{center}
\end{figure}

Denoting  $x$ such that $T_x =\pm X$, the resolutions Eq.~(\ref{E_res}) have a counterpart at the level of edges in $E$,
\begin{equation}\label{E_res}
x = a_1\oplus (a_2 \oplus (a_3 \oplus ... (a_{n-1}\oplus a_n))),\; a_k\in E_{\cal{O}},\; \forall k=1,..,n.
\end{equation}
The resolution Eq.~(\ref{E_res}) has an interpretation in cohomological terms. Namely, we have the following property.
\begin{Lemma}\label{GeomResol}
For every inferable observable $X$, $\pm T_x =X \in {\cal{I}}$,  there exists a 2-chain $f\in C_2(E)$ and a 1-chain $e\in C_1(E)$ with $\{e\} \subset E_{\cal{O}}$, such that $x+e =\partial f$.
\end{Lemma}
Therein, we have used the shorthand $\{e\}$ for the set of edges $a\in E$ that appear in the expansion of the 1-chain $e\in C_1(E)$. See Fig.~\ref{lemon} for illustration.\smallskip

{\em{Proof of Lemma~\ref{GeomResol}.}}  (a) Main case: $X\not\in {\cal{O}}$. With Definition~\ref{Infer} and Eq.~(\ref{E_res}), we may express the resolution of $x$ as a set of nested equalities $x=a_1\oplus b_1$, $b_1= a_2 \oplus b_2$, ... , $b_{n-2} = a_{n-1}+b_{n-1}$, $b_{n-1}=a_n$, where $a_1,..,a_n \in E_{\cal{O}}$ and $b_1,.., b_{n-2} \in E$. By the definition of $\cC(E)$, for every elementary step of inference there is a corresponding face in $F$. With Eq.~(\ref{partial}), the above nested relations can be rewritten as
$$
\begin{array}{rcl}
x+ a_1 + b_1 &=&\partial f_1,\\
b_1 + a_2 + b_2 &=& \partial f_2,\\
\multicolumn{3}{c}{...}\\
b_{n-3} + a_{n-2} + b_{n-2} &=& \partial f_{n-2},\\
b_{n-2}+a_{n-1}+ a_n &=& \partial f_{n-1},
\end{array}
$$
where $f_1,f_2,..., f_{n-1}\in F$. Adding those equations and setting $f:=\sum_{i=1}^{n-1}f_i$, we obtain $x+e = \partial f$, and $f\in C_2(E)$.

(b) Degenerate case: $X \in {\cal{O}}$. We may represent $x$ by the resolution $x=x\oplus 0$, where $0$ is such that $T_0=I$. The corresponding 2-chain is $f=[x|0]$. $\Box$

\subsection{The input group}\label{iG}

At the center of the present generalization is the notion of the {\em{input group}} $Q$. In all MBQC, the classical input enters into the computation through the choice of the measurement setting. In $H$-MBQC, the set of possible inputs is described by a group, $Q$. For every input $q\in Q$, the corresponding measurement setting is obtained from a reference setting corresponding to the trivial input $q=\text{e}$.

In any given $H$-MBQC, we denote by ${\cal{E}}_\text{e} \subset E_{\cal{O}}$ the set of edges corresponding to the observables measured when the input is trivial, $q=\text{e}$. ${\cal{E}}_\text{e}$ is such that it gives rise to a resolution Eq.~(\ref{E_res}). The resolutions for other inputs $q\in Q$ are obtained via
$$
{\cal{E}}_q = q({\cal{E}}_\text{e}),
$$
The computational procedure for converting an input $q\in Q$ to an output $o(q)$ is then described by Procedure I displayed in Table~\ref{Proc1}.

For Procedure I to be executable and consistent with our notion of inference, i.e., the resolution, the following constraints have to be satisfied. 
\begin{enumerate}
\item{$Q$ preserves $E_{\cal{O}}$,
\begin{equation}\label{Qc1}
q a \in E_{\cal{O}},\; \forall a\in E_{\cal{O}},\, \forall q\in Q.
\end{equation}}
\item{$Q$ is compatible with the addition ``$\oplus$''. I.e., for all $a,b \in E$ such that $[T_a,T_b]=0$, and $a\in {\cal{O}}$ or $b\in {\cal{O}}$ it holds that
\begin{equation}\label{Qc2}
q(a\oplus b) = qa \oplus qb,\; \forall q\in Q.
\end{equation}}
\end{enumerate}

\begin{table}
\begin{center}
\begin{tabular}{l}
\textbf{Procedure I}\\ \hline \hline MBQC with input $q \in Q$\\ \hline
\parbox{0.85\textwidth}{
\begin{enumerate}
\item{{\em{Classical pre-processing.}} The elements $a_1,.., a_n \in {\cal{E}}_\text{e}$, with ${\cal{E}}_\text{e}$ a resolution Eq.~(\ref{E_res}),
are transformed according to
$$
a_i \mapsto q a_i.
$$}
\item{{\em{Measurement.}} The observables $T_{q a_i}=\eta(q a_i)$, for $i=1,..,n$ are measured on the resource quantum state $|\Psi\rangle$, and the corresponding measurement outcomes $s(q a_i) \in \mathbb{Z}_2$ are recorded.}
\item{{\em{Classical post-processing.}} The computational output is obtained via 
$$
o(q) = \sum_{i=1}^n s(q a_i) \mod 2.
$$}
\end{enumerate}}\\ \hline
\end{tabular}
\caption{\label{Proc1}Procedure I, the standard procedure for running an $H$-MBQC.} 
\end{center}
\end{table}

Property Eq.~(\ref{Qc1}) ensures that if the $T_{a_i} \in {\cal{O}}$ (input $q=\text{e}$), then $T_{qa_i} \in {\cal{O}}$ for all $q\in Q$. Hence, Step 2 of the above Procedure I can be executed for all inputs $q\in Q$. Property Eq.~(\ref{Qc2}) ensures that the observables producing the computational output in Step 3 of Procedure I form a resolution, and hence the inference is valid.\smallskip

We can now adjust the definition of the set $E_0$ to the computational setting of $H$-MBQC. First, denote by $a_\text{e} \in E$ the edge defined by the resolution
\begin{equation}\label{aeDef}
a_\text{e}:=a_1\oplus (a_2 \oplus ... \oplus (a_{n-1} \oplus a_n)), \; \text{with }\{a_1,..,a_n\} = {\cal{E}}_\text{e}.
\end{equation}
Then,
\begin{equation}\label{E0Def}
E_0:=\{qa_\text{e},\, q\in Q\}.
\end{equation}
As we shall see, this definition of $E_0$ works both for deterministic and probabilistic $H$-MBQC. This completes the adaption of the definitions of ${\cal{I}}$, ``$\oplus$'' and $E_0$  to the present computational setting.\medskip

We remark that $l2$-MBQC is a special case of $H$-MBQC. The notion of an input group is already present therein, although not emphasized. In $l2$-MBQC 
the inputs form an Abelian group $(\mathbb{Z}_2)^m$, for some integer $m$. Namely, the classical pre-processing relation Eq.~(\ref{CCR_in}) is rewritten in terms of the action of the input group $Q$ on the set $E_{\cal{O}}$ as follows. The elements $a\in E_{\cal{O}}$ have two labels, the site $i$ and the label $q_i\in \mathbb{Z}_2$ representing the choice of measurement basis. We thus write them as $a_{i,q_i}$. The generator $\alpha_k$ of $Q$ corresponds to the $k$-th column of the matrix $S$ in Eq.~(\ref{CCR_in}), and the action of the generators $\alpha_k$ on $E$ reproducing Eq.~(\ref{CCR_in}) is
$$
\alpha_k: 
\begin{array}{rl}   
a_{i,0} \circlearrowright,  \, a_{i,1} \circlearrowright,& \text{if }S_{ik}=0,\\
a_{i,0}\longleftrightarrow a_{i,1}, & \text{if }S_{ik}=1.
\end{array}
$$
It is easily checked that this is a group action on $E_{\cal{O}}$. \medskip

\stepcounter{expl}{\em{Example, Part~\theexpl.}} The input relation Eq.~(\ref{CCR_inGHZ}) is represented by the action of a group $Q$ on a reference context. In the present GHZ-MBQC, the reference context is ${\cal{E}}_{\text{e}}=(a_{X_1},a_{X_2},a_{X_3})$, and the input group is $Q=\mathbb{Z}_2\times \mathbb{Z}_2$. Its two generators $q_1,q_2$ are related to the input $x$, $y$ of the OR-gate via
\begin{equation}\label{InpMap}
x \mapsto q_1,\; y \mapsto q_2,
\end{equation}
and act on the edges in $E$ via
\begin{equation}\label{Qghz}
\begin{array}{rl}
q_1: &a_{X_1} \leftrightarrow a_{Y_1}, \;  a_{X_3} \leftrightarrow a_{Y_3}, \; a_{X_2}  \circlearrowright,\; a_{Y_2} \circlearrowright,\\
q_2: &a_{X_2} \leftrightarrow a_{Y_2}, \;  a_{X_3} \leftrightarrow a_{Y_3}, \; a_{X_1}  \circlearrowright,\; a_{Y_1} \circlearrowright.
\end{array}
\end{equation}
We may verify that this action of the input group reproduces  Eq.~(\ref{CCR_inGHZ}) for the GHZ-MBQC. For example, if $a=b=0$ then $q=\text{e}$, and the observables to be measured are $X_1$, $X_2$ and $X_3$, in accordance with Eq.~(\ref{CCR_inGHZ}). Further, if $a=1$ and $b=0$, then with Eq.~(\ref{InpMap}) the corresponding input group element is $q_1$, and with Eq.~(\ref{Qghz}) it follows that ${\cal{E}}_{q_1} = (a_{Y_1},a_{X_2},a_{Y_3})$. Hence, $Y_1$, $X_2$ and $Y_3$ are measured, in agreement with Eq.~(\ref{CCR_inGHZ}). The other two cases are analogous.

\subsection{The symmetry group}\label{sG}

Each $H$-MBQC comes with a symmetry group $H$. In this section, (i) we describe the action of the symmetry group $H$ on the set ${\cal{I}}$ of observables, and (ii) its action on the $H$-MBQCs themselves.

Denote by ${\cal{R}}_o$ the set of operators related to computational output,
\begin{equation}\label{Rdef}
{\cal{R}}_o :=  \left\{ (-1)^{o(q)}\prod_{a \in {\cal{E}}_\text{e}} T_{qa}, q\in Q\right\}.
\end{equation}
Recall that therein ${\cal{E}}_\text{e}$ is the set of edges corresponding to the observables in the measurement sequence for the trivial input $q=\text{e}$. The order of operators $T$ in each  product is the same as in the corresponding measurement sequence. Note that the operators in ${\cal{R}}_o$ comprise the observables measured in $H$-MBQC for all inputs $q$, cf. Step 2 of Procedure I, and the output function. \smallskip

The symmetry group $H$  satisfies the following conditions.
\begin{enumerate}
\item{$H$ preserves ${\cal{O}}$, i.e., 
\begin{equation}
\label{PresM}
h(A) \in {\cal{O}},\; \forall A \in {\cal{O}},\forall h \in H.
\end{equation}}
\item{$H$ preserves all Abelian subgroups of ${\cal{I}}$, i.e., 
\begin{equation}
\label{Autom}
h(A)h(B) = h(AB),\; \forall A,B \in {\cal{I}}\, \text{with}\, [A,B]=0,\forall h \in H.
\end{equation}}
\item{The set of operators ${\cal{R}}_0$ is preserved by $H$,
\begin{equation}\label{oPres}
h({\cal{R}}_o) = {\cal{R}}_o,\;\;\forall h \in H.
\end{equation}}
\end{enumerate}
Eq.~(\ref{oPres}) relates the output function $o$ and the action of the symmetry $H$ on ${\cal{O}}$. As we will discuss further below, in the deterministic case this constrains the symmetry, and in the probabilistic case it constrains the output function.\medskip

\stepcounter{expl}{\em{Example, Part~\theexpl.}} For the GHZ-MBQC, the symmetry group is
\begin{equation}\label{uH}
H= \langle A_1A_2Y_3, A_1Y_2A_3,Y_1A_2A_3\rangle,
\end{equation} 
where $A:=(X+Y)/\sqrt{2}$, and the action of the symmetry group elements $h\in H$ on the observables $T\in {\cal{O}}$ via $h(T):=h Th^\dagger$. We have ${\cal{R}}_o=\{X_1X_2X_3, -X_1Y_2Y_3, -Y_1X_2Y_3,-Y_1Y_2X_3 \}$ in this example, and the invariance condition Eq.~(\ref{oPres}) is easily verified. For example, for $h_0:=A_1A_2Y_3$ we find
$$
h_0:\; X_1X_2X_3 \longleftrightarrow  -Y_1Y_2X_3,\;  -X_1Y_2Y_3 \longleftrightarrow -Y_1X_2Y_3.
$$
\noindent
\textbf{Action of $H$ on $H$-MBQCs.} Next we discuss the action of the symmetry transformations $h \in H$ on a given $H$-MBQC. Essentially, for every MBQC ${\cal{M}}$, in the transformed MBQC $h({\cal{M}})$ the observable $h(T_a)$ is measured whenever $T_a$ is measured in ${\cal{M}}$. The outcome $s'(a)$ for the observable $h(T_a)$ replaces the outcome $s(a)$ for $T_a$ in obtaining the computational output. 

More formally, for all transformations $h \in H$, the MBQC according to Procedure I, but transformed under $h$, is the Procedure II displayed in Table~\ref{Proc2}.

\begin{table}
\begin{center}
\begin{tabular}{l}
\textbf{Procedure II} \\ \hline \hline MBQC with input $q \in Q$, transformed under $h \in H$\\ \hline 
\parbox{0.85\textwidth}{
\begin{enumerate}
\item{{\em{Classical pre-processing.}} The edges $a_1,.., a_n \in E_{\cal{O}}$, with ${\cal{E}}_\text{e}$ a resolution Eq.~(\ref{E_res}), are transformed according to
$$
a_i \mapsto q a_i.
$$}
\item{{\em{Measurement.}} The observables $h(T_{q a_i})$, for $i=1,..,n$ are measured on the resource quantum state $|\Psi\rangle$, and the corresponding measurement outcomes $s'(q a_i) \in \mathbb{Z}_2$ are recorded.}
\item{{\em{Classical post-processing.}} The computational output is obtained via 
$$
o'(q) = \sum_{i=1}^n s'(q a_i) \mod 2.
$$}
\end{enumerate}}\\ \hline
\end{tabular}
\caption{\label{Proc2}Procedure II, the standard procedure acted on by a symmetry transformation  $h\in H$.}
\end{center}
\end{table}

As a consequence of the above definitions, the symmetry transformations $H$ preserve the corresponding $H$-MBQC in the following sense.
\begin{Theorem}\label{Invar}
Given an $H$-MBQC ${\cal{M}}$ with a symmetry group $H$, for all $h \in H$, $h({\cal{M}})$ compute the same function $o$ with the same average probability of success.
\end{Theorem}
It is because of this result that we call $H$ the symmetry group of an MBQC. 

{\em{Proof of Theorem~\ref{Invar}.}}  Related to the set ${\cal{R}}_o$ of Eq.~(\ref{oPres}) we define an operator 
\begin{equation}\label{WitDef}
{\cal{W}}_o := \sum_{q\in Q} (-1)^{o(q)}\prod_{a \in {\cal{E}}_\text{e}} T_{qa} = \sum_{A \in {\cal{R}}_o} A.
\end{equation}
With Eq.~(\ref{oPres}) it thus follows that
\begin{equation}\label{WitInvar}
h({\cal{W}}_o) = {\cal{W}}_o,\; \forall h \in H.
\end{equation}
In an $H$-MBQC ${\cal{M}}$, the probability $P_o(q,\rho)$ for the measured output to match the intended output $o(q)$ in Procedure I, given the input $q\in Q$ and resource state $\rho$, is
$$
P_o(q,\rho) = \left\langle \frac{I +(-1)^{o(q)}\prod_{a\in {\cal{E}}_\text{e}} T_{qa}}{2} \right\rangle_\rho.
$$
Now with Eq.~(\ref{WitDef}), the success probability $\overline{P}_o(\rho)$ of the $H$-MBQC, averaged with equal weight over all inputs $q\in Q$, is
\begin{equation}\label{AvP_S}
\overline{P}_o(\rho) = \frac{1}{2} + \frac{\langle {\cal{W}}_o\rangle_\rho}{2|Q|}.
\end{equation}
Correspondingly, for any $H$-MBQC $h({\cal{M}})$, performed according to Procedure II with given input $q\in Q$ and resource state $\rho$, the probability $P^{(h)}_o(q,\rho)$ for the measured output to match $o(q)$ is $P^{(h)}_o(q,\rho) = 1/2 +(-1)^{o(q)}  \left\langle \prod_{a\in {\cal{E}}_\text{e}} h(T_{qa})\right\rangle_\rho/2$. The average success probability of $h({\cal{M}})$ computing $o$ therefore is
$$
\overline{P}^{(h)}_o(\rho) = \frac{1}{2} + \frac{\langle h({\cal{W}}_o)\rangle_\rho}{2|Q|}.
$$
Comparing the above expressions for $\overline{P}_o(\rho)$ and $\overline{P}^{(h)}_o(\rho)$, with the invariance condition Eq.~(\ref{WitInvar}) it holds that $\overline{P}^{(h)}_o(\rho) = \overline{P}_o(\rho)$, for all $h\in H$. $\Box$

\subsection{Symmetry and input}\label{SI}

In this section we show that the two groups introduced in Sections~\ref{iG} and \ref{sG}, the input group and the symmetry group, are related. Namely, as we now show, the input group $Q$ is obtained from the symmetry group $H$ by modding out a certain normal subgroup.

The symmetry group $H$ has a normal subgroup $N$ defined by the property that for all $n \in N$ it holds that $n a = a$, $\forall a \in E$.  $H/N$ inherits an action on $E$ from $H$, through the relation $h(T_a) = \pm T_{ha}$, cf. Eq.~(\ref{PhiTilDef}). By definition of $N$, $(hn)a=h(na) = ha$, and therefore, with $[h]$ the equivalence class representing $h\in H$ in the quotient $H/N$, it holds that $ha = [h]a$. 

We have the following relation between $H$ and the input group $Q$.
\begin{Theorem}\label{quot}
For an MBQC with symmetry group $H$, the input group $Q$ may always be chosen
\begin{equation}\label{Qdef}
Q = H/N.
\end{equation}
\end{Theorem}
{\em{Proof of Theorem~\ref{quot}.}} We need to show that the group $Q$ defined through Eq.~(\ref{Qdef}) satisfies Eqs.~(\ref{Qc1}) and (\ref{Qc2}). Consider any $h\in H$. Since $h$ satisfies Eq.~(\ref{PresM}), by the above induced action $[h] \in H/N$ satisfies Eq.~(\ref{Qc1}). Further, choose $a,b$ such that $a\oplus b$ is defined. Then $T_{a\oplus b}=\pm T_aT_b$, and by Eq.~(\ref{Autom}), $h(T_{a\oplus b})=\pm h(T_a)h(T_b)$. Therefore, $[h](a\oplus b) = [h]a \oplus [h]b$ which is Eq.~(\ref{Qc2}). $\Box$\medskip

\stepcounter{expl}{\em{Example, Part~\theexpl.}} Here we verify Theorem~\ref{quot} for the GHZ-MBQC. Inspecting the generators of the symmetry group $H$ in Eq.~(\ref{uH}) and the generators of $Q$ in Eq.~(\ref{Qghz}), we find that, when acting on $E$,
$$
A_1A_2Y_3 \mapsto q_1q_2,\; A_1Y_2A_3 \mapsto q_1,\; Y_1A_2A_3 \mapsto q_2.
$$
The images of the generators of $H$ are dependent, and generate the group $Q=\mathbb{Z}_2\times \mathbb{Z}_2$ as they should. The normal subgroup $N$ of $H$ consists of Pauli operators, which do not change any edge labels $a\in E$.\medskip

\subsection{Physical meaning of cohomological invariance}

Consider an observable $T_a \in {\cal{O}}$. Flipping it into $-T_a$ makes the corresponding $H$-MBQC neither easier nor harder to perform. Further, flipping an observable $T_b \in {\cal{I}}\backslash {\cal{O}}$ has no effect on the computation at all. Therefore, changing the function $\eta$ defined in Eq.~(\ref{etaDef}),
\begin{equation}\label{GauCha}
\eta(a) = T_a \longrightarrow \eta'(a) = (-1)^{\gamma(a)}T_a,\; \; \text{for any } \gamma: E\longrightarrow \mathbb{Z}_2,
\end{equation}
should be considered as an equivalence transformation, or change of gauge.

By Eqs.~(\ref{prod_T}) an (\ref{PhiTilDef}), the corresponding changes in the cocycles $\beta$ and $\tilde{\Phi}$ are
\begin{equation}\label{GauCha2}
\beta \longrightarrow \beta + d \gamma,\;\;\tilde{\Phi} \longrightarrow \tilde{\Phi} + \tt{d} \gamma.
\end{equation}
It is thus the cohomologically invariants $[\beta]$ and $[\tilde{\Phi}]$ which contain the physical information that is unaffected by the gauge changes of Eq.~(\ref{GauCha}).

\section{Deterministic $H$-MBQC}\label{comp}

Here we discuss deterministic $H$-MBQCs. We establish the triple $\mathfrak{S}=(\cC(E,E_0),\beta_\Psi, Q)$ as an algebraic structure describing MBQC, satisfying the Criteria (I) and (II). Namely, $\beta_\Psi$ contains the computational output $o: Q\longrightarrow \mathbb{Z}_2$, and the action of $Q$ on $\cC(E,E_0)$ can be used to address the output $o(q)$ for any $q\in Q$. Further, there exists a natural equivalence relation among output functions $o$, and the corresponding equivalence classes only depend on $[\beta_\Psi]$; hence they are a cohomological invariant. Finally, $[\beta_\Psi]$ is also a conextuality witness.

\subsection{Computational output}\label{OutBeta}

\begin{figure}
\begin{center}
\begin{tabular}{lcl}
(a) & \mbox{ } & (b) \vspace{2mm}\\
\includegraphics[width=4cm]{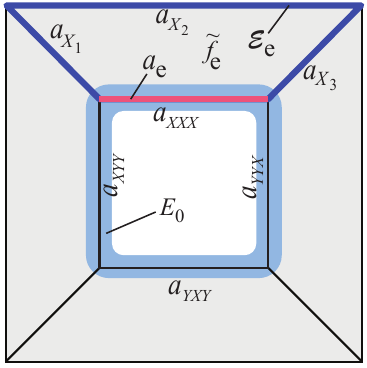} && \includegraphics[width=4cm]{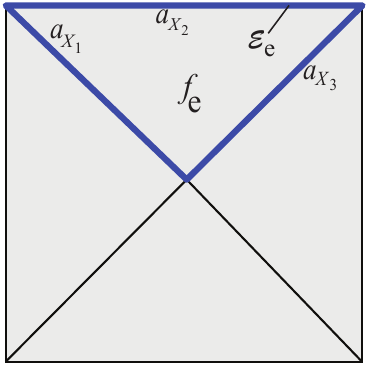}
\end{tabular}
\caption{\label{Illu}The chain complexes $\cC(E)$ and $\cC(E,E_0)$ for the example of GHZ-MBQC. (a) The complex $\cC(E)$. Shown are the edge $a_\text{e}$, the face $\tilde{f}_\text{e}$, and the sets ${\cal{E}}_\text{e}$ and $E_0$. The edge labels correspond to the observables in Fig.~\ref{MermSt}\,b. (b) The complex $\cC(E,E_0)$ obtained from $\cC(E)$ by contracting the edges in $E_0$. Highlighted is the face $f_\text{e}$, which is the image of $\tilde{f}_\text{e}$ under the contraction.}
\end{center}
\end{figure}

Recall the definition of $a_\text{e}$ in Eq.~(\ref{aeDef}), and that ${\cal{E}}_\text{e}$ is the ordered set of edges in $E_{\cal{O}}$ corresponding to the observables measured for the trivial input $q=\text{e}$. By Lemma~\ref{GeomResol} there exists a 2-chain $\tilde{f}_\text{e} \in C_2(E)$ such that
\begin{equation}\label{feTilDef}
\partial \tilde{f}_\text{e} = a_{\text{e}}+\sum_{a\in {\cal{E}}_\text{e}}a.
\end{equation} 
Furthermore, $f_\text{e}$ is the corresponding face in $\cC(E,E_0)$, obtained from $\tilde{f}_\text{e}$ by contraction of the edge $a_\text{e}$. Hence, $\partial_R f_\text{e} = \sum_{a\in {\cal{E}}_\text{e}}a$.
The objects ${\cal{E}}_\text{e}$, $a_\text{e}$ and $f_\text{e}$ are illustrated for the GHZ-example in Fig.~\ref{Illu}.

Denoting the function evaluation $q\mapsto \beta_\Psi(q\partial_R f_\text{e})$ as $\beta_\Psi(\,\cdot \, \partial_R f_\text{e})$, we have the following result.

\begin{Theorem}\label{PC}
For every $H$-MBQC with input group $Q$ there is a 2-chain $f_\text{e}\in C_2(E,E_0)$ such that
\begin{equation}\label{Hcomp}
o(\,\cdot \,)  = \beta_\Psi(\,\cdot \, f_{\text{e}}).
\end{equation}
\end{Theorem}
{\em{Proof of Theorem~\ref{PC}.}} Acting by $q$ on Eq.~(\ref{feTilDef}), we obtain
$\partial \,q \tilde{f}_\text{e} = qa_{\text{e}}+\sum_{a\in {\cal{E}}_\text{e}} qa,\; \forall q\in Q$. Therefore, for the measurement $s$ obtained in the measurement sequence for the input $q$,
$$
ds (q\partial \tilde{f}_e) =s_\Psi(qa_\text{e}) + \sum_{a\in {\cal{E}}_\text{e}} s(qa).
$$
From Eq.~(\ref{prod_T}) it follows that $ds(q\partial \tilde{f}_e)=\beta(q\partial \tilde{f}_e)$. Further recall Eq.~(\ref{betaPsiDef}), $\beta_\Psi(q f_\text{e}) = \beta(q\tilde{f}_\text{e})+s_\Psi(q a_\text{e})$. Thus,
$$
\beta_\Psi(q f_\text{e}) =  \sum_{a\in {\cal{E}}_\text{e}} s(qa) = o(q).
$$
Therein, in the second equality we used the measurement-outcome-to-output relation in Step 3 of Procedure I. $\Box$\medskip

The conclusion is that the cocycle $\beta_\Psi$, together with the action of the input group $Q$ on the complex $\cC(E,E_0)$, specifies the computational output $o: Q \longrightarrow \mathbb{Z}_2$.

\subsection{Equivalent output functions}\label{EqO}

Under the gauge change Eq.~(\ref{GauCha}) the cocycle $\beta$ changes, and hence also $\beta_\Psi$,
$$
\beta_\Psi \longrightarrow \beta_\Psi + d \gamma|_{E\backslash E_0}.
$$
In turn, the outputted function $o$ also changes. Yet, these changes of gauge do not change the hardness of performing $H$-MBQC, and we therefore group group output functions $o$ that result from another by change of gauge into equivalence classes $[o]$. 
\begin{Cor}\label{Cor1}
Consider two $H$-MBQC which only differ in their corresponding 2-cocycles, $\beta_\Psi$ vs. $\beta_\Psi'$. If $[\beta_\Psi] = [\beta_\Psi']$ then $[o]=[o']$. 
\end{Cor}
\smallskip

The conclusion is that the physically motivated equivalence classes $[o]$ of output functions depend only on the cohomology classes $[\beta_\Psi]$; hence are cohomological invariants.\smallskip

\stepcounter{expl}{\em{Example, Part~\theexpl.}} In the GHZ-MBQC, we may flip  $Y_3 \longrightarrow - Y_3$. In result, the new computed function is an AND. Therefore, AND and OR are equivalent wrt. MBQC. Considering the whole set of equivalence transformations for this example, we find that there are two equivalence classes of functions on two bits, the non-linear Boolean functions and the linear ones. Each member of the former class  boosts the classical control computer of MBQC to computational universality, whereas the second class has no effect on the computational power at all. 

From the cohomological perspective, $H^2({\cal{C}}(E,E_0),\mathbb{Z}_2) = \mathbb{Z}_2$ in the GHZ case, i.e. there are two equivalences classes of cocycles $\beta_\Psi$. The trivial class corresponds to the linear Boolean functions on two bits and the non-trivial class to the non-linear Boolean functions.

\subsection{Contextuality}\label{context}

The cohomology class $[\beta_\Psi]$ is a contextuality witness. This result has been established in \cite{Coho}, and we restate it here to complement the findings of Sections~\ref{OutBeta} and \ref{EqO}.

\begin{Theorem}[\cite{Coho}]\label{Thm_beta}
If for a deterministic $H$-MBQC ${\cal{M}}$ it holds that $[\beta_\Psi]\neq0 \in H^2(\cC(E,E_0),\mathbb{Z}_2)$, then ${\cal{M}}$ is contextual.
\end{Theorem}
{\em{Proof of Theorem~\ref{Thm_beta}.}}
Assume that the $H$-MBQC ${\cal{M}}$ is non-contextual, and that therefore at least one value assignment $\sm: E \longrightarrow \mathbb{Z}_2$ satisfying $\sm|_{E_0}=s_\Psi$ exists. Now let  $s= \sm + \overline{s}_\Psi$. Thus, $s|_{E_0} = 0$, and hence $s \in C^1(E,E_0)$. Further, $ds = d\sm + d\overline{s}_\Psi = \beta + d\overline{s}_\Psi = \beta_\Psi$, and thus $[\beta_\Psi]=0$. $\Box$

\section{Probabilistic $H$-MBQC}\label{ProbComp}

Most quantum algorithms known to date are probabilistic. It is therefore of interest to obtain a probabilistic extension of the results presented in Section~\ref{comp}. Probabilistic computation occurs when the observables corresponding to the output don't stabilize the resource state, for example when the resource state is mixed. 

At the center of the probabilistic scenario lurks a serious puzzle for the fundamental computational structures $\mathfrak{S}$ which the present paper aims to identify.  Namely, in any given MBQC with input group $Q$, for {\em{every}} function $o:Q \longrightarrow \mathbb{Z}_2$ there is a success probability $\overline{P}_o$ with which, averaged over all inputs $q \in Q$, this function is computed. For some functions, this success probability may be high, and for others low. But we can no longer say that a specific function is being computed while all others are not. 

For example, in the GHZ-MBQC, the OR-function is computed with certainty. However, as soon as probabilistic computations are admitted, we may as well say that the same physical setup computes the constant function $o\equiv 1$ with a $75\,\%$ success probability.  

From this perspective, the outputted function is merely a matter of interpretation, a parameter that can be freely chosen.  This poses a problem for the notion of computational structure $\mathfrak{S}$ as constrained by Criterion (I). If the outputted function $o$ is only a free parameter, then what is the meaning of Criterion (I), namely that the fundamental computational structure $\mathfrak{S}$ contains $o$?

Here, symmetry comes to the rescue. Namely, the invariance condition Eq.~(\ref{oPres}), $H(\mathcal{R}_o)=\mathcal{R}_o$, constrains the output function $o$, since the definition of $\mathcal{R}_o$ depends on it. Closer inspection shows that the output function $o$ is determined by symmetry up to an additive constant.\medskip

Three of the four subjects we treat in this section are direct counterparts of the corresponding subjects in Section~\ref{comp}; namely Criterion (I), Criterion (II), and the cohomological description of these. In Section~\ref{COiH} we prove the statement made above, namely that symmetry constrains the output function up to a constant. In Section~\ref{eqoP} we demonstrate that a physically motivated equivalence class of output functions is a cohomological invariant.  In Section~\ref{ProbCon} we discuss contextuality in probabilistic $H$-MBQCs, emphasizing the role of cohomological invariants. And then there is a fourth subject: The discussion in this section is based on a cocycle in group cohomology, i.e., it is different from the cocycle $\beta_\Psi$ used in Section~\ref{comp}. In Section~\ref{compat_P} we show that the cohomological description of the present probabilistic scenario is nonetheless compatible with the cohomological description of the deterministic scenario in the previous section.

\subsection{Computational output}\label{COiH}

In this section we investigate the constraints on the output function $o$ imposed by the symmetry condition Eq.~(\ref{oPres}).

Consider $h({\cal{R}}_o)$, for any $h\in H$. With the action of $h$ on ${\cal{R}}_o$ specified element-wise  by Eq.~(\ref{PhiTilDef}), and denoting by $[h]$ the equivalence class of $h$ in $Q=H/N$, we have
$$
h({\cal{R}}_o) = \left\{(-1)^{o(q)} \prod_{a\in {\cal{E}}_\text{e}} h(T_{qa}), \,q\in Q \right\}=  \left\{ (-1)^{o(q)+\sum_{a\in {\cal{E}}_\text{e}} \tilde{\Phi}_h(qa) } \prod_{a\in {\cal{E}}_\text{e}} T_{[h]qa}, \, q\in Q\right\}.
$$
Requiring therein the invariance property Eq.~(\ref{oPres}), we find
\begin{equation}\label{oPhiRel1}
o([h]q)= o(q) + \tilde{\Phi}_h(q\partial_R f_\text{e}).
\end{equation}
As we detail below, this  constraint on the output function $o$ by the phase function $\tilde{\Phi}$ determines the output function up to an additive constant.

However, first we reformulate Eq.~(\ref{oPhiRel1}) to better suit our cohomological description. Presently, the constraint on $o$ is expressed in terms of the phase function $\tilde{\Phi}$, which is a group cocycle of the symmetry group $H$, and evaluates on the chain complex $\cC(E)$. We aim to re-express the constraint in terms of a new phase function $\Phi$, yet to be derived from the existing p hase function $\tilde{\Phi}$, which (a) evaluates on the chain complex $\cC(E,E_0)$ or a substructure thereof, for consistency with the deterministic case; (b) is defined on the input group $Q=H/N$ rather than the symmetry group $H$, for better match with the output function $o$ which is also defined on $Q$; and (c) which is a group coycle wrt. the group $Q$, to preserve the cohomological flavour of the constraint.

To derive the new phase function $\Phi$ from $\tilde{\Phi}$, first note that the r.h.s. of Eq.~(\ref{oPhiRel1}) only requires evaluations of $\tilde{\Phi}$ on edges in $\cC(E,E_0)$. We consider $\tilde{\Phi}$ in a suitable gauge, $\Phi_\chi:= \tilde{\Phi}+\tt{d}\overline{\chi}$, where $\overline{\chi}|_E = \chi$, and $\overline{\chi}: E \longrightarrow \mathbb{Z}_2$ is identically zero on $E\backslash E_0$.  Thus, the change of gauge does not affect Eq.~(\ref{oPhiRel1}). $\Phi_\chi$ is a 1-cycle in group cohomology, 
\begin{equation}
\label{1cocycle}
{\tt{d}}\Phi_\chi =0,
\end{equation}
since ${\tt{d}}\tilde{\Phi}=0$  and $\tt{d}\tt{d} \overline{\chi}=0$. Further, $\chi$ can always be chosen such that
\begin{equation}\label{PhiChiProp}
(\Phi_\chi)_h(a) =0,\; \forall a\in E_0,\,\forall h \in H,
\end{equation}
hence $\Phi_\chi$ is a 1-cocycle of $H$ acting on $\cC(E,E_0)$. 

For our applications, $\Phi_\chi$ is restricted to act on 1-chains in $\cC_Q$. Eq.~(\ref{oPhiRel1}) becomes
\begin{equation}\label{oPhiRel1b}
o([h]q)= o(q) + (\Phi_\chi)_h(q\partial_R f_\text{e}),
\end{equation}
with $\Phi_\chi\in C_1(H,{\cal{C}}_Q)$, and ${\tt{d}}\Phi_\chi=0$.

We observe that the l.h.s. of Eq.~(\ref{oPhiRel1b}) depends only on $[h] \in Q=H/N$ whereas the r.h.s. formally depends on $h$. Consistency requires that $\tilde{\Phi}_h(q\partial_R f_\text{e})$ depends on $h$ only through $[h]$. We now prove this property. Specializing $h\in H$ to $n\in N$ in Eq.~(\ref{oPhiRel1b}) yields $(\Phi_\chi)_n(q\partial_R f_\text{e})=0$, $\forall q\in Q$, $n \in N$. Employing Eq.~(\ref{CG2}), we obtain $(\Phi_\chi)_{nh}(q\partial_R f_\text{e})=(\Phi_\chi)_h(q\partial_R f_\text{e})$, $\forall h\in H$, $\forall n\in N$.

Thus we can define
\begin{equation}\label{DefPhi}
\Phi_{[h]}(x):=(\Phi_\chi)_h(x),\;\forall [h]\in Q,\,\forall x\in C_1(\cC_Q),
\end{equation}
where $\cC_Q$ is a sub-complex of $\cC(E,E_0)$, with edge set $E({\cal{C}}_Q)$ and face set $F(\cC_Q)$ given by
\begin{equation}
\label{EFC}
E(\cC_Q) = \{q \partial_R f_\text{e},\,q\in Q\},\;\; F(\cC_Q)=\{q f_\text{e},\,q\in Q\}.
\end{equation}
The phase function $\Phi \in C^1(Q,{\cal{C}}_Q)$ is the quantity of interest. By construction, it satisfies the properties (a) and (b) required above, namely that it evaluates on (a substructure of) $\cC(E,E_0)$ and is defined on $Q$ rather than $H$.

We still need to verify property (c), namely that $\Phi$ is a 1-cocycle with respect to the group $Q$. For all $x\in C_1(\cC_Q)$ and all $[h_1],[h_2]\in Q$, and recalling that all addition is mod 2, we have
$$
\begin{array}{rcl}
{\tt{d}} \Phi([h_1],[h_2],x) &=& \Phi([h_1],[h_2]x)+\Phi([h_1][h_2],x)+ \Phi([h_2],x)\\
&=& \Phi_\chi(h_1,h_2x)+\Phi_\chi(h_1h_2,x)+\Phi_\chi(h_2,x)\\
&=& {\tt{d}} \Phi_\chi(h_1,h_2,x)\\
&=& 0.
\end{array}
$$
Above, in the first and third line we have used the definition Eq.~(\ref{CG2}) of the coboundary of 1-chains in group cohomology, in the second line Eq.~(\ref{DefPhi}), and in the fourth line with Eq.~(\ref{1cocycle}). Thus, $\Phi$ is a 1-cocycle in the cohomology of the group $Q$, as required.

We have the following result.
\begin{Theorem}\label{o_Inv}
For any $H$-MBQC ${\cal{M}}$, with symmetry group $H$ and corresponding input group $Q$, the output function $o:Q\longrightarrow \mathbb{Z}_2$ is specified by the symmetry condition Eq.~(\ref{oPres}) up to an additive constant,
\begin{equation}\label{oPhiRel2}
o(q) = o(\text{e}) + \Phi_q(\partial_R f_\text{e}),\;\; \forall q\in Q.
\end{equation}
\end{Theorem}
Thus, the group cocycle $\Phi$ defined in Eq.~(\ref{DefPhi}) determines the output function $o$ almost completely.\medskip

{\em{Proof of Theorem~\ref{o_Inv}.}} Using the definition of $\Phi$ in Eq.~(\ref{DefPhi}), we rewrite Eq.~(\ref{oPhiRel1b}) as
$$
o([h]q)= o(q) + \Phi_{[h]}(q\partial_R f_\text{e}).
$$
Now specializing to $q=\text{e}$, and subsequently relabeling $[h] \longrightarrow q$ in the above yields Eq.~(\ref{oPhiRel2}). $\Box$

\subsection{Equivalent output functions}\label{eqoP}

For the probabilistic case, we consider a slightly broader notion of equivalence than in the previous deterministic case. As before, two output functions are equivalent if they are related by a change of gauge Eq.~(\ref{GauCha}). Additionally, we also we consider them equivalent if they differ by a constant offset $c\in \mathbb{Z}_2$. This reflects the fact that, in the probabilistic scenario, the phase function determines the output only up to an additive constant, cf. Eq.~(\ref{oPhiRel2}). To further motivate this additional equivalence, note that the flipping-over of the measurement bases corresponding to the gauge change Eq.~(\ref{GauCha}) can be modelled as not flipping the measurement bases, but adding an offset 1 to the measurement outcomes. If such an offset can be added to individual measurement outcomes before the classical side processing, a similar constant offset may be added to the output $o$ after that processing. Thus, in the probabilistic scenario, a physically motivated notion for equivalence classes $[[o]]$ of output functions is 
$$
[[o]] =\left\{ o + \gamma (\,\cdot\, \partial_R f_\text{e}) + c,\;\forall \gamma: E\backslash E_0\longrightarrow \mathbb{Z}_2,\, \forall c\in \mathbb{Z}_2 \right\}.
$$
We then have the following result as a corollary to Theorem~\ref{o_Inv}.
\begin{Cor}\label{Cor2}
Consider two $H$-MBQCs with given input group $Q$ that differ in the their respective phase function $\Phi$ vs. $\Phi'$. If $[\Phi] = [\Phi'] \in H^1(Q,{\cal{C}}_R)$ then $[[o]]=[[o']]$.\end{Cor}
{\em{Proof of Corollary~\ref{Cor2}.}} The assumption of the corollary is that $\Phi'=\Phi +{\tt{d}}\gamma$, for some $\gamma: E\backslash E_0 \longrightarrow \mathbb{Z}_2$. Therefore, with Eq.~(\ref{oPhiRel2}), it holds that $o'(q)+o'(\text{e}) = \Phi_q(\partial_R f_\text{e}) + ({\tt{d}}\gamma)_q(\partial_R f_\text{e})= \Phi_q(\partial_R f_\text{e})+\gamma(q\partial_R f_\text{e})+ \gamma(\partial_R f_\text{e})$. Thus, again with Eq.~(\ref{oPhiRel2}),
$$
o'(q) = o(q) + \gamma(q\partial_R f_\text{e}) + c,
$$
where $c:=o(\text{e})+o'(\text{e})+\gamma(\partial_R f_\text{e})$. Hence, $o$ and $o'$ are in the same equivalence class $[[o]]$. $\Box$

\subsection{Compatibility between the probabilistic and the deterministic case}\label{compat_P}

Section~\ref{comp} provides a separate treatment of the deterministic case which is somewhat simpler than that of the general probabilistic case, as it is based on the cocycle $\beta_\Psi$ rather than a phase function $\Phi$. Yet, probabilistic $H$-MBQC contains the deterministic case as a limit, and we therefore need to check consistency between the two treatments. 

In deterministic $H$-MBQC, the output function $o$ is fully specified by the cocycle $\beta_\Psi$, whereas in the general probabilistic case, the phase function $\Phi$ determines $o$ only up to an additive constant, cf. Theorem~\ref{o_Inv}. From Theorem~\ref{PC} we can derive the weaker prediction, as it is made in the general probabilistic case, and state it for the deterministic case. It is
\begin{equation}\label{obet}
o(q) + o(\text{e}) =  \beta_\Psi(q f_\text{e}) + \beta_\Psi(f_\text{e}),\; \forall q\in Q.
\end{equation}
This has to agree with the statement of Theorem~\ref{o_Inv}. For consistency we thus require $\beta_\Psi(q f_\text{e}) + \beta_\Psi(f_\text{e}) =\Phi_q(\partial_R f_\text{e})$. In this regard, we note the following relation.
\begin{Lemma}[\cite{Coho2}]\label{BetPhiLem}
For any $\chi:E_0 \longrightarrow \mathbb{Z}_2$, the phase function $\Phi_\chi:=\tilde{\Phi} +{\tt{d}}\overline{\chi}$ and the cocycle $\beta_\chi:=\beta + d\overline{\chi}$ are related via
\begin{equation}
d\Phi_\chi = {\tt{d}}\beta_\chi.
\end{equation}
\end{Lemma}
{\em{Proof of Lemma~\ref{BetPhiLem}.}} First, we prove the corresponding result in $\cC(E)$,
\begin{equation}\label{betaPhi}
d\tilde{\Phi} = \ttd \beta.
\end{equation}
Namely, with Eq.~(\ref{Autom}) we have, for any $h\in H$,
$$
(-1)^{\tilde{\Phi}_h(a\oplus b)}T_h(a\oplus b) =h(T_{a\oplus b}) = (-1)^{\beta(a,b)} h(T_a)h(T_b) =  (-1)^{\beta(a,b)+\tilde{\Phi}_h(a)+\tilde{\Phi}_h(b)+\beta(ha,hb)} T_{h(a\oplus b)}.
$$
Thus, ${\tt{d}}\beta\, (h,[a|b])=\tilde{\Phi}(h,\partial [a|b])$, which is Eq.~(\ref{betaPhi}) applied to $(h,[a,b])$. This proves Eq.~(\ref{betaPhi}). 

Now, $d\Phi_\chi= d\tilde{\Phi} + d{\tt{d}}\overline{\chi} = {\tt{d}}\beta  + d{\tt{d}}\overline{\chi}= {\tt{d}}(\beta+ d \overline{\chi}) = {\tt{d}} \beta_\chi$, where we have used Eq.~(\ref{betaPhi}) and $d{\tt{d}}={\tt{d}}d$. $\Box$\smallskip

Applying Lemma~\ref{BetPhiLem} to the setting at hand yields
\begin{equation}\label{dbdP2}
\beta_\chi([h] f_\text{e})+\beta_\chi(f_\text{e}) = (\Phi_\chi)_h(\partial_R f_\text{e}) =  (\Phi_\chi)_{[h]}(\partial_R f_\text{e}) = \Phi_{[h]}(\partial_R f_\text{e}),
\end{equation}
where for the last equality we have used Eq.~(\ref{DefPhi}). 

It remains to show that, in the deterministic case, $\chi \equiv s_\Psi$.  We note that the observables in the set ${\cal{R}}_o$ defined in Eq.~(\ref{Rdef}) are all proportional to some corresponding $T_x$, for $x\in E_0$, i.e.,
$$
{\cal{R}}_o = \left\{ (-1)^{\chi(x)} T_x,\; x\in E_0\right\}.
$$ 
Comparing with the original definition of ${\cal{R}}_o$ in Eq.~(\ref{Rdef}), we find
\begin{equation}\label{chiobet}
\chi(qa_\text{e}) = o(q)+\beta(q\tilde{f}_\text{e}).
\end{equation}
Therein, $a_\text{e}\in E_0$ represents the observable (up to sign) corresponding to the input $q=\text{e}$, and $\tilde{f}_e$ is the face in $\cC(E)$ which, upon contraction of the edges $E_0$ becomes the face $f_\text{e}$ in $\cC(E,E_0)$. When applying Eq.~(\ref{chiobet}) to the deterministic case, we find that this agrees with Theorem~\ref{PC} if and only if
$$
\chi(x) = s_\Psi(x), \;\forall x\in E_0\;\;\; \text{(in the deterministic case)}.
$$
Inserting this into Eq.~(\ref{dbdP2}) yields $\beta_\Psi(qf_\text{e})+\beta_\Psi(f_\text{e}) = \Phi_q(\partial_R f_\text{e})$, for all $q\in Q$. Eq.~(\ref{obet}) thus agrees with Theorem~\ref{o_Inv}, as required. 

\subsection{Contextuality}\label{ProbCon}

Here we provide a version of Theorem~\ref{NLPCrel} suited to our cohomological description. We identify a threshold in the success probability of $H$-MBQC beyond which the computation is contextual, and show that this threshold is a cohomological invariant. We remark that the two theorems presented in this section have close counterparts in \cite{Coho2}; cf. Section~5.3 therein. The difference is that the present results are based on the computational output $o: Q\longrightarrow \mathbb{Z}_2$, whereas the results in \cite{Coho2} are based on a partial value assignment $\chi:E_0 \longrightarrow \mathbb{Z}_2$. It is conceptually more transparent to derive the results of this section from scratch.\smallskip

Be $\Lambda_{\cal{O}}$ the set of Boolean functions on $E_{\cal{O}}$, $\Lambda_{\cal{O}} := \set{s: E_{\cal{O}} \longrightarrow \mathbb{Z}_2}$, and $\H(o,o')$ the Hamming distance between two functions $o,o': Q\longrightarrow \mathbb{Z}_2$. Based on this we define 
\begin{equation}\label{HamDef}
\H^*(o,\Lambda_{\cal{O}}):= \min_{c\in \mathbb{Z}_2} \min_{s\in \Lambda_{\cal{O}}} \H(o(\cdot)+c, s(\,\cdot\, \partial_R f_\text{e})).
\end{equation}
From Eq.~(\ref{oPhiRel2}) we know that the output function $o$ depends on the phase function $\Phi$ and a constant offset; hence the same holds for the Hamming distance $\H^*$. The minimization over $c\in \mathbb{Z}_2$ in Eq.~(\ref{HamDef}) cancels the dependence on the offset, such that $\H^*$ is a function of $\Phi$ only. In fact, as we prove below, $\H^*$ depends only on the cohomology class $[\Phi]$. Before this, we turn to the phenomenological significance of $\H^*$.
\begin{Theorem}\label{Tctx}
An $H$-MBQC ${\cal{M}}$ with resource state $\rho$, input group $Q$, output function $o$ and average success probability $\overline{P}_o(\rho)$ is contextual  if 
\begin{equation}\label{ncie2}
\overline{P}_{o}(\rho) > \displaystyle{1 -\frac{\H^*(o,\Lambda_{\cal{O}})}{|Q|}}.
\end{equation}
\end{Theorem}
This theorem is an adaption of Corollary~3 in \cite{Coho2} to our present computational setting.\smallskip 

\stepcounter{expl}{\em{Example, Part~\theexpl.}} In the 3-qubit MBQC, we have $\mathbb{H}^*(\text{OR},\Lambda_{\cal{O}})=1$. Therefore, the contextuality threshold according to Theorem~\ref{Tctx} is $\overline{P}_\text{OR}=0.75$, in agreement with the Mermin inequality.\medskip

{\em{Proof of Theorem~\ref{Tctx}.}} Assume an ncHVM with value assignments $\Lambda$ and a probability distribution $q$. The ncHVM expression $\overline{P}_{o}(p)$ for the quantity $\overline{P}_{o}(\rho)$ satisfies
$$
\begin{array}{rcl}
\overline{P}_{o}(p) &=& \displaystyle{\frac{1}{|Q|} \sum_{\sm \in \Lambda, q \in Q} q(\sm) \delta_{o(q),\sm(q\partial_R f_\text{e})}} \\
&\leq& \displaystyle{ \frac{1}{|Q|} \max_{\sm\in \Lambda} \sum_{q\in Q} \delta_{o(q),\sm(q\partial_R f_\text{e})}} \\
&=& \displaystyle{ \frac{1}{|Q|} \max_{\sm\in \Lambda} \sum_{q\in Q} \delta_{o(q),\sm|_{E_{\cal{O}}}(q\partial_R f_\text{e})}} \\
&\leq& \displaystyle{ \frac{1}{|Q|} \max_{s \in \Lambda_{\cal{O}}} \sum_{q\in Q} \delta_{o(q),s(q\partial_R f_\text{e})}} \\
&\leq& \displaystyle{ \frac{1}{|Q|} \max_{c\in\mathbb{Z}_2}\max_{s \in \Lambda_{\cal{O}}} \sum_{q\in Q} \delta_{o(q)+c,s(q\partial_R f_\text{e})}} \\
&=& \displaystyle{ 1-\frac{\H^*(o,\Lambda_{\cal{O}})}{|Q|}.}
\end{array}
$$  
Herein, in the fourth line we have used that for all $\mathfrak{s}\in \Lambda$ it holds that $\mathfrak{s}|_{E_{\cal{O}}} \in \Lambda_{\cal{O}}$. We thus find that if $\overline{P}_{o}(\rho)$ is larger than $1 - \H^*(o,\Lambda_{\cal{O}})/|Q|$ then no ncHVM can account for that. $\Box$\medskip
 
From Eq.~(\ref{oPhiRel2}) and the definition Eq.~(\ref{HamDef}) we know that the contextuality threshold Eq.~(\ref{ncie2}) is a property of the group cocycle $\Phi$. However, we can say more.
\begin{Theorem}\label{Hamming2}
Let ${\cal{M}}$ be an $H$-MBQC with input group $Q$ and phase function $\Phi$ as defined in Eq.~(\ref{DefPhi}). Then, the contextuality threshold Eq.~(\ref{ncie2}) depends on the group cocycle $\Phi$ only through the group cocycle class $[\Phi]$, i.e., it is a cohomological invariant. 
\end{Theorem}

{\em{Proof of Theorem~\ref{Hamming2}.}} We recall that the phase function $\Phi$, defined in Eq.~(\ref{DefPhi}), is obtained from the phase function $\Phi_\chi$ through restriction of the application domain $\cC(E,E_0)$ to $\cC_Q$ defined in Eq.~(\ref{EFC}), and correspondingly of $H$ to $Q=N/N$. Therefore, for any change of gauge $\gamma$ transforming $\Phi \mapsto \Phi + {\tt{d}}\gamma$ there is a corresponding $\tilde{\gamma}$ transforming $\Phi_\chi \mapsto \Phi_\chi + {\tt{d}}\tilde{\gamma}$. 

With Eq.~(\ref{oPhiRel2}), the change in the output function affected by a gauge change $\gamma$ is $o \mapsto o'(\cdot) = o(\cdot) +{\tt{d}}\gamma(\cdot, \partial_R f_\text{e})$. Therefore, 
$$
\begin{array}{rcl}
\H^*(o',\Lambda_{\cal{O}}) 
&=& \min_{c\in \mathbb{Z}_2}\min_{s \in \Lambda_{\cal{O}}}\H(o(\cdot)+\gamma(\cdot\, \partial_Rf_\text{e})+\gamma(\partial_R f_\text{e})+c,s(\cdot\, \partial_R f_\text{e}))\\
&=& \min_{c\in \mathbb{Z}_2}\min_{s \in \Lambda_{\cal{O}}}\H(o(\cdot)+c,s(\cdot\, \partial_R f_\text{e})+\gamma(\cdot\, \partial_Rf_\text{e}))\\
&=& \min_{c\in \mathbb{Z}_2}\min_{s \in \Lambda_{\cal{O}}}\H(o(\cdot)+c,s(\cdot\, \partial_R f_\text{e})+\tilde{\gamma}(\cdot\, \partial_Rf_\text{e}))\\
&=& \min_{c\in \mathbb{Z}_2}\min_{s \in \Lambda_{\cal{O}}}\H(o(\cdot)+c,\left(s+\tilde{\gamma}|_{E_{\cal{O}}}\right)(\cdot\, \partial_Rf_\text{e}))\\
&=& \min_{c\in \mathbb{Z}_2}\min_{s \in \Lambda_{\cal{O}}}\H(o(\cdot)+c,s(\cdot\, \partial_Rf_\text{e}))\\
&=& \H^*(o,\Lambda_{\cal{O}}).
\end{array}
$$
Therein, in the second line we have absorbed $\gamma(\partial_R f_\text{e})$ in the constant $c$ over which the expression is minimized, and $\gamma(\cdot\, \partial_Rf_\text{e})$ is moved to the other side. In the third line we have used the remark made at the beginning of this proof. In the fourth line we used the fact that, by construction, for all $q\in Q$ the relative 1-cycle $q \partial_Rf_\text{e}$ runs inside $E_{\cal{O}}$. In the fifth line we used $s\in \Lambda_{\cal{O}} \Longleftrightarrow s+\tilde{\gamma}|_{E_{\cal{O}}} \in \Lambda_{\cal{O}}$.

We have shown that the Hamming distance $\H^*(o,\Lambda_{\cal{O}})$ is unaffected by all gauge changes $\Phi\mapsto \Phi + {\tt{d}}\gamma$. Since the contextuality threshold depends on $o$ only through $\H^*(o,\Lambda_{\cal{O}})$, it is invariant. $\Box$
%\newpage
 
\section{Conclusion}\label{concl}

In this paper we have demonstrated that quantum computations are amenable to symmetry analysis, similar in spirit to particles in high energy physics and phases in condensed matter physics. To this end, we have described a variant of measurement-based quantum computation, called $H$-MBQC, in which each computation is acted on by a corresponding symmetry group $H$ that leaves the computational output invariant. 

In this scheme of quantum computation, the computational output is a function $o:Q \longrightarrow \mathbb{Z}_2$, where the input domain is a group $Q$. The following two properties hold: (a) The input group $Q$ is derived from the symmetry group $H$, namely $Q=H/N$ with $N$ the normal subgroup of $H$ that maps each measurable observable to itself up to sign (Theorem~\ref{quot}); and (b) The action of $H$ on the measurable observables defines a 1-cocycle $\Phi$ in the cohomology of the input group $Q$, which determines the output function $o$ up to an additive constant (Theorem~\ref{o_Inv}).

Now focussing on the general  probabilistic case, $H$-MBQC has the following cohomological features:
\begin{enumerate}
\item{The physically motivated equivalence classes $[[o]]$ of output functions depend on $[\Phi]$ only, i.e., are cohomological invariants (Corollary~\ref{Cor2}).}
\item{The threshold success probability beyond which an $H$-MBQC becomes contextual is a function of $[\Phi]$ only, hence a cohomological invariant (Theorems~\ref{Tctx},\ref{Hamming2}).}
\end{enumerate}
Thus, with regard to the initially posed Criteria (I) and (II), the essential information about $H$-MBQC is of cohomological nature. 

We note that the above results have been established only for the limiting case of $H$-MBQCs with flat temporal order. The general case of $H$-MBQC with proper temporal order remains for future work.\smallskip

The focus of this work are {\em{macroscopic}} properties of quantum computations, namely the computational output and the presence of contextuality, both of which characterize the quantum computation as a whole. We remark that there is a complimentary line of inquiry which, again from the vantage point of symmetry, looks at {\em{microscopic}} properties of quantum computation. Namely it investigates which quantum gates are guaranteed by the presence of a given symmetry in a corresponding symmetry-protected phase \cite{SPT1}--\cite{SPT5}. The macroscopic and the microscopic approach have two elements in common, one physical and one mathematical. The physical commonality is that they are both concerned with measurement-based quantum computation, and the mathematical commonality is that they both employ group cohomology. It is conceivable that these two approaches look at opposite ends of the same structure, and may be unified in a larger framework.\bigskip

\noindent
{\em{Acknowledgments.}} I thank C. Okay and E. Tyhurst for discussions, and acknowledge support from NSERC.\smallskip

This paper is dedicated to Dr. Klaus Weidig, my mother Marina Rau{\ss}endorf, Dr. Manfred Gubsch, RA Uwe Wunderlich, and Matthias Kluge, and to the memory of Karl Friedrich. They kept a ship afloat, and Klaus Weidig rebuilt it.

\end{document}